\documentclass[12pt,english]{article}
\pdfoutput=1

\usepackage[a4paper,lmargin=70pt,rmargin=70pt]{geometry}
\usepackage{setspace}
\usepackage{babel}
\usepackage{graphicx}
\usepackage{amsmath}
\usepackage{amssymb}
\usepackage[colorlinks=true,urlcolor=blue,anchorcolor=blue,citecolor=blue,filecolor=blue,linkcolor=blue,menucolor=blue,pagecolor=blue,linktocpage=true]{hyperref}

\usepackage{cite}

\newcommand{\ed}{\,.}
\newcommand{\ec}{\,,}
\newcommand{\ecq}{\ec\quad}

\newcommand{\bZ}{\ensuremath{\mathbb{Z}}}

\newcommand{\cO}{\ensuremath{\mathcal{O}}}

\newcommand{\barh}{\bar{h}}
\newcommand{\bartau}{\bar{\tau}}

\DeclareMathOperator{\trace}{Tr}

 \newcommand{\es}[2] {\begin{equation} \label{#1} \begin{split} #2 \end{split} \end{equation}}

\numberwithin{equation}{section}

\begin{document}

\begin{flushright} 
\today\\
SU-ITP-xx/xx
\end{flushright} 

\vspace{0.1cm}

\begin{center}
  {\LARGE 2D CFT Partition Functions at Late Times}
\end{center}
\vspace{0.1cm}
\vspace{0.1cm}

\begin{center}
Ethan D{\sc yer} and Guy G{\sc ur-Ari}

\vspace{0.3cm}

  {\it Stanford Institute for Theoretical Physics,\\
Stanford University, Stanford, CA 94305, USA}
\end{center}

\onehalfspacing

\begin{center}
  {\bf Abstract}
\end{center}

We consider the late time behavior of the analytically continued partition function $Z(\beta + it) Z(\beta - it)$ in holographic $2d$ CFTs.
This is a probe of information loss in such theories and in their holographic duals.
We show that each Virasoro character decays in time, and so information is not restored at the level of individual characters.
We identify a universal decaying contribution at late times, and conjecture that it describes the behavior of generic chaotic $2d$ CFTs out to times that are exponentially large in the central charge.
It was recently suggested that at sufficiently late times one expects a crossover to random matrix behavior.
We estimate an upper bound on the crossover time, which suggests that the decay is followed by a parametrically long period of late time growth.
Finally, we discuss integrable theories and show how information is restored at late times by a series of characters.
This hints at a possible bulk mechanism, where information is restored by an infinite sum over non-perturbative saddles.

\newpage
\tableofcontents{}
\newpage

\section{Introduction and Summary}
\label{sec:intro}

Quantum black holes have finite entropy and a discrete spectrum of states.
The details of this spectrum are inaccessible in the semi-classical approximation: The density of states one obtains from the Bekenstein-Hawking entropy is a smooth function of the energy.
In this work we address the question of how the discrete spectrum arises in $2d$ conformal field theories and their holographic duals.

Maldacena suggested that one may address this question by studying the late time behavior of correlation functions \cite{Maldacena:2001kr}, which is a sharp probe of the discrete energy levels in the spectrum.
For unitary systems with discrete spectra, connected thermal correlators of the form $\langle O(t) O(0) \rangle$ (where $O$ is a Hermitian operator) tend to decay exponentially until times of order the entropy $S$, and then proceed to oscillate erratically about zero with an RMS amplitude of order $e^{-S}$.\footnote{
  By the notation $e^{-S}$ we mean that the quantity scales as $e^{-n_{\rm dof}}$ where $n_{\rm dof}$ is the number of degrees of freedom.
    We will be interested in $2d$ CFTs with large central charge $c$, for which $n_{\rm dof} \sim c$.
  We note that at very late times of order $e^{e^S}$ we expect recurrences, which do not play a role in this work.
  See \cite{Barbon:2003aq,Barbon:2004ce,Cardy:2014rqa} for a discussion of recurrences in the context of information loss.
  }
On the other hand, correlation functions computed in a classical black hole background tend to decay exponentially forever.
This decay is often referred to as `information loss'.

Holography may be a useful setting for studying the question of how a discrete black hole spectrum arises.
From the boundary field theory point view, the fact that the spectrum is discrete is trivial if we place the theory on a compact spatial manifold.
Similarly, the qualitative features of the late time behavior follow easily from mild assumptions about the spectrum (such as the fact that the theory is chaotic).
The challenge is then to describe this behavior in `bulk language', using objects that are natural from a gravity point of view.
In this work we focus on another quantity that is also sensitive to information loss at late times.

\subsection{Spectral Form Factor and Information Loss}

Consider the thermal partition function $Z(\beta)$, and let us analytically continue $\beta \to \beta + it$.
The parameter $t$ should be thought of as real time.
Let $E_n$ be the discrete energy levels, each with degeneracy $N_n$, and consider the following quantity.
\begin{align}
  g(\beta,t) \equiv |Z(\beta+it)|^2 =
  \sum_{n,m} N_n N_m e^{-\beta(E_n + E_m) + it(E_n - E_m)} \,.
  \label{g0}
\end{align}
If we formally set $\beta=0$ then \eqref{g0} becomes a well-studied quantity in Quantum Chaos called the \textit{spectral form factor} (for reviews, see \cite{Guhr:1997ve,fyodorov2005introduction}).
We will use the same name to refer to $g(\beta,t)$ at any $\beta$.
In the context of black hole physics this quantity was first discussed in \cite{Papadodimas:2015xma}, and was recently studied in the context of information loss in the Sachdev-Ye-Kitaev model \cite{SY,K} in \cite{Cotler2016}.
See also \cite{Cardy:2014rqa} for a related discussion.

At late times the double sum in \eqref{g0} essentially localizes onto terms with $E_n = E_m$.
As we review in Section~\ref{sec:sff}, the time average of $g(\beta,t)$ obeys the bound
\begin{align}
  \bar{g}(\beta) \equiv
  \lim_{t_o \to \infty} \frac{1}{t_o} \int_0^{t_o} g(\beta,t) dt \ge
  Z(2\beta) \ed \label{gbound0}
\end{align}
The bound is saturated when the spectrum has no degeneracies.
The non-zero time average reflects a weighted counting of the discrete energy levels in the spectrum.
The quantity on the right-hand side is of order $e^S$.

On the other hand, suppose we have a bulk theory with a black hole background and focus on the BTZ black hole for simplicity.
We approximate the exact partition function by the BTZ black hole partition function $Z(\beta) = \exp\left(\frac{\pi^2 c}{3\beta}\right)$, which is the dominant contribution for temperatures above the Hawking-Page transition.
We then find a spectral form factor that decays to 1 at late times.
If we also include the 1-loop determinant we find that the spectral form factor decays to zero, representing no discrete states in the corresponding spectrum.\footnote{
  That the BTZ partition function decays to 1 (if we do not include the 1-loop determinant) is related to the fact that the inverse Laplace transform of $e^{1/\beta}$ is given by $E^{-1/2} I_1(2E^{1/2}) + \delta(E)$ which includes a single discrete state.
  }
We see that the spectral form factor, just like the correlation function, is sensitive to information loss.
See \cite{Festuccia:2006sa,Iizuka:2008hg} for related discussions.

It was suggested in \cite{Maldacena:2001kr} that one may improve the situation by adding subleading bulk saddle points such as thermal AdS$_3$.
It is easy to check that including the thermal AdS$_3$ contribution indeed raises the time average, but this contribution is not sufficient for the time average to obey the bound \eqref{gbound0} at high temperature.
Indeed, we will see that no finite number of subleading saddles is enough to obey the bound \eqref{gbound0} at high temperature.

For $2d$ conformal field theories, the question of information loss in the thermal two point function was studied in \cite{Kleban:2004rx} and for collapsing black holes in \cite{Anous:2016kss}. Recently, the authors of \cite{Fitzpatrick:2014vua, Fitzpatrick:2015zha, Fitzpatrick:2016ive,Fitzpatrick:2016mjq} considered the four point function of two heavy operators $O_H$, $\Delta_{H}\sim c$, and two light operators $O_L$ on the cylinder $\langle O_{H}|O_{L}(\phi,it)O_{L}(0)|O_{H}\rangle$.
This is a microcanonical version of the calculation described above.
In the large $c$ limit, corresponding to the classical black hole limit in the bulk, one finds that the correlation function reproduces the thermal two-point function on a line, with temperature set by the heavy operator's dimension and thus decays in time.
In \cite{Fitzpatrick:2016ive,Fitzpatrick:2016mjq} it was speculated that perhaps the late time decay is avoided (and information is restored) within each Virasoro block in an OPE expansion of this four-point function.
This question is difficult to answer because the relevant Virasoro blocks are not known exactly.

We are able to answer this question in our context, by considering instead the spectral form factor, which has a decomposition in terms of Virasoro characters, analogous to the Virasoro blocks that show up in the OPE expansion of the heavy-heavy-light-light correlator.\footnote{
  The torus partition function can be written as a correlator involving 4 heavy twist operators.
  The Virasoro characters are the blocks that appear in an OPE expansion of this correlator.
  }
  The Virasoro characters have known closed-form expressions, and each relevant Virasoro character decays at late times.
  We conclude that in chaotic $2d$ theories information is not restored kinematically in general, namely as a consequence of Virasoro symmetry, but rather dynamically, due to an interplay between infinitely many characters.
  (Integrable theories will be discussed separately.
  For such theories information loss still occurs at the level of Virasoro characters, but is explicitly restored in the characters of the extended chiral algebra.)

The authors of \cite{Maloney:2007ud,Keller:2014xba} studied the discrete spectrum of chaotic $2d$ CFTs by working directly with the thermal partition function.
Our conclusion agrees with their results.
The authors of \cite{Maloney:2007ud} considered a modular invariant partition function that is made up of the vacuum character plus its modular images (appropriately regulated).
They found that the corresponding density of states is essentially smooth and captures almost none of the discrete states.
Here we advertise that if one is interested only in whether or not the spectrum contains discrete states (rather than in the detailed properties of these states), it is enough to check whether the time-averaged spectral form factor $\bar{g}(\beta)$ vanishes, a potentially simpler computation.

This discussion of information loss has been phrased in terms of the boundary Virasoro characters, but also has a natural bulk interpretation.
The character which dominates at high temperature corresponds to the bulk BTZ saddle.
The $\cO(1/c)$ correction to the character corresponds to a one-loop determinant in the bulk.
Therefore, a resolution of information loss phrased in terms of Virasoro characters would probably shed light on how information is restored in the bulk.

\subsection{Late Times and Random Matrix Theory}
\label{sec:rmt}

The late time behavior of the spectral form factor is only sensitive to the structure of small energy differences.
We generally expect that if we probe any chaotic system at sufficiently small energy differences, then the Hamiltonian can be approximated by a random matrix chosen from a suitable Gaussian ensemble.
The authors of \cite{Cotler2016} made the observation that the late time behavior of chaotic theories should therefore be described by random matrix theory (see \cite{Guhr:1997ve,fyodorov2005introduction} for a review of RMT).
This was verified for the Sachdev-Ye-Kitaev model \cite{SY,K} in \cite{You:2016ldz}. We thus now turn to RMT as a guide for what to expect for the late time behavior of the spectral form factor.

\begin{figure}
  \centering
    \includegraphics[width=0.6\textwidth]{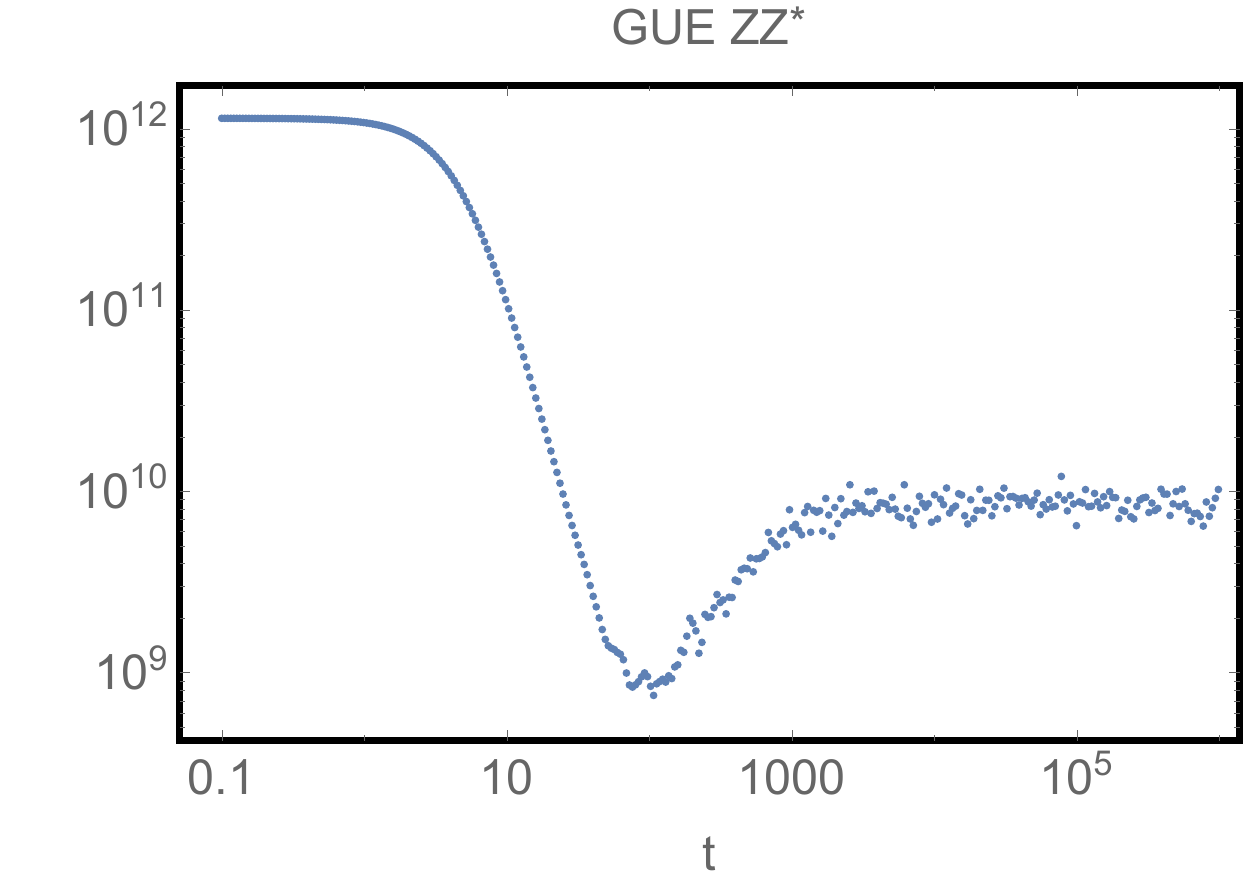}
    \caption{The spectral form factor $g(\beta,t)$ for the GUE ensemble of random matrices, using 50 matrices of rank 2,000 and computed with $\beta=5$.
      At late times we have a period of close to linear growth we call the ramp, folowed by a plateau.
    }
\label{fig:GUEg}
\end{figure}
Figure~\ref{fig:GUEg} shows the spectral form factor for random matrices selected from the Gaussian Unitary Ensemble (GUE).
We will discuss this curve in more detail below.
For now we merely point out that (i) the shape of the curve before its minimum (the dip) is dominated by the coarse-grained shape of the spectrum (in the case of Gaussian random matrices this is Wigner's semicircle law), and that (ii) after the dip time the curve starts probing the discrete energy levels.
In particular, the period of linear growth is related to the spectral rigidity of random matrix energy levels (essentially the fact that energy levels repel).

In \cite{Cotler2016} it was conjectured that the existence of a dip time, followed by a period of linear growth, is a generic feature of chaotic systems, including black holes.
The value of the dip time is non-universal and depends on the detailed properties of the theory, including the coarse-grained shape of the spectrum that determines the early time decay.
Here we test this conjecture in the context of $2d$ CFTs, and find evidence that the approximate dip time is robust for chaotic CFTs dual to gravity.

\begin{figure}
  \centering
    \includegraphics[width=0.6\textwidth]{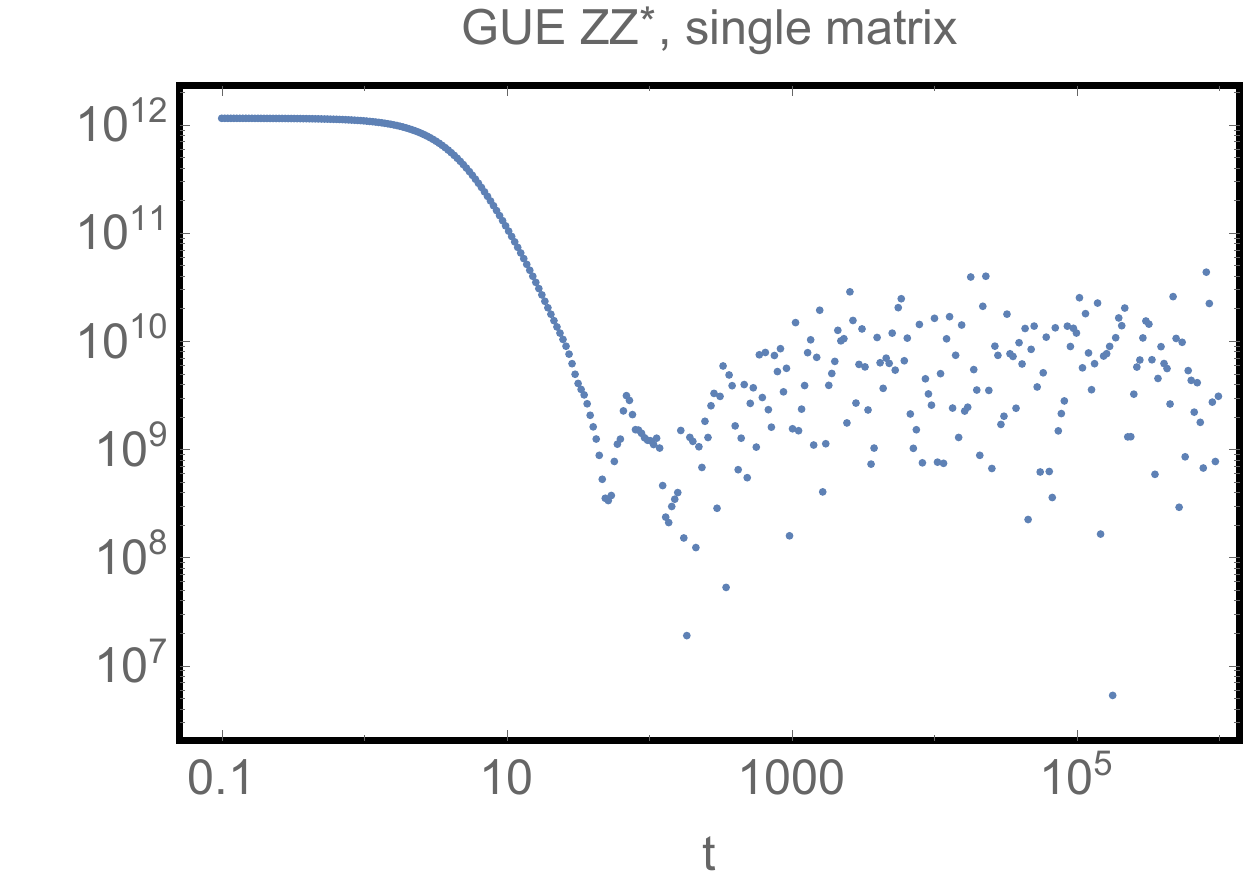}
    \caption{The spectral form factor $g(\beta,t)$ for a single matrix, using the same parameters as Figure~\ref{fig:GUEg}.
      The late time ramp and plateau are barely visible.
    }
\label{fig:GUEgSingle}
\end{figure}
In RMT (and in the SYK model) the spectral form factor is defined by averaging over an ensemble of Hamiltonians.
This averaging leads to a smooth curve at late times.
In trying to apply the conjecture to an ordinary quantum field theory, one has to confront the fact that there is only one Hamiltonian.
As a result, the late time behavior is expected to be erratic.
Figure~\ref{fig:GUEgSingle} shows the spectral form factor computed from a single GUE matrix.
Beyond the dip time the fluctuations become large and the features of Figure~\ref{fig:GUEg} are barely visible (see also \cite{prange1997spectral}).
However, as explained in \cite{Cotler2016}, one can replace ensemble averaging by time averaging over a parametrically small window (in the limit of a large Hilbert space dimension), restoring the late time features.
It is therefore meaningful to discuss the random matrix theory ramp and plateau at late times even in an ordinary quantum field theory.

In this work we estimate the dip time at which a generic $2d$ CFT crosses over into the RMT regime.
This is done by estimating the shape of the early decay of the curve using modular invariance, and assuming that at late times we have the linear growth predicted by RMT.
Our estimate relies on identifying the dominant contribution from a single Virasoro character at each point in time.
We estimate that the spectral form factor decays at late times in an erratic way, with an envelope that decays as
\begin{align}
  g(\beta,t) \lesssim \exp \! \left( \frac{\pi^2 c}{3\beta} \right)
  \cdot \frac{1}{t} 
  \ed
\end{align}
We will see this implies a parametrically long period of linear growth following the dip time.
We expect the same to be true of black holes in AdS$_3$.

\subsection{Summary of Results}

Here is a brief summary of the key points of this paper.
\begin{enumerate}
\item The late time behavior of the spectral form factor $|Z(\beta+it)|^2$ is a useful probe of the discreteness of the spectrum \cite{Cotler2016}.
  As in the case of two-point functions, a decay at late times indicates that we are not probing the discrete states of the spectrum, and signals information loss.
\item In $2d$ CFTs the partition function has an expansion in terms of the Virasoro characters.
  Each character decays at late times, and therefore Virasoro symmetry is not enough in general to restore information.
\item In a chaotic theory we expect the late time behavior to be described by random matrix theory.
  In particular, we expect there to be a characteristic time scale (the dip time) beyond which the RMT description is valid.
\item We estimate the dip time in $2d$ CFTs.
 Although this contribution is not controlled for all time, we conjecture a dip time which scales as $e^c$.
  Beyond the dip time we expect there to be a period of linear growth (with large fluctuations) that is parameterically long at large $c$ and high temperature.
\item For certain integrable models, or BPS sub-sectors of generic models, we identify a precise infinite set of bulk saddles which restore the information naively lost in the leading thermodynamic approximation.
\end{enumerate}

The rest of the paper is organized as follows.
In Section~\ref{sec:sff} we discuss the spectral form factor and information loss in $2d$ CFTs.
In Section~\ref{sec:2d} we review the Virasoro character expansion and the modular properties of the torus partition function, and provide simple estimates of its decay before the dip time.
Then, in Section~\ref{sec:uni} we give an improved estimate of the decay by identifying the dominant character at any rational time.
We conclude that these contributions are not sufficient to avoid information loss.
In Section~\ref{sec:dip} we estimate the dip time, beyond which we expect the system to have an effective random matrix theory description.
In Section~\ref{sec:int} we discuss integrable theories.
We show that for certain integrable theories, or BPS sectors of generic theories, information is restored by identifying the dominant saddle point at each particular time using modular invariance.
Appendix~\ref{app:bh} gives a short review of black holes in AdS$_3$.

\section{Spectral Form Factor}
\label{sec:sff}

In this section we define the spectral form factor and discuss its properties in relation to information loss.
Consider a unitary quantum field theory with a holographic dual.
Place the theory on a compact manifold so that it has a discrete spectrum.
The spectrum consists of energy levels $E_n$, each with degeneracy $N_n$.
The density of states is given by
\begin{align}
  \rho(E) = \sum_n N_n \delta(E-E_n) \ed \label{rhoExact}
\end{align}
The thermal partition function at inverse temperature $\beta$ is 
\begin{align}
  Z(\beta) \equiv \trace \left( e^{-\beta H} \right) 
  = \sum_n N_n e^{-\beta E_n} \ed \label{Z}
\end{align}
We assume for simplicity that this function is finite for any $\beta > 0$ (this is always true for $2d$ CFTs).
Let us generalize the partition function and define
\begin{align}
  Z(\beta + it) \equiv \trace \left( e^{-\beta H - i H t} \right)
  = \sum_n N_n e^{-\beta E_n - i E_n t} \ed
\end{align}
One can obtain this function by analytically continuing $Z(\beta)$, taking $\beta \to \beta + it$.
The parameter $t$ is conveniently thought of as real time.
We then define the \textit{spectral form factor} by
\begin{align}
  g(\beta,t) \equiv \left| Z(\beta + it) \right|^2
  = \sum_{n,m} N_n N_m e^{-\beta(E_n + E_m) + it(E_n - E_m)}
  \ed
\end{align}
This is an important quantity in the study of random matrix theory \cite{Guhr:1997ve,fyodorov2005introduction}.

In this work we will study the late time behavior of $g(\beta,t)$.
In a general chaotic theory this behavior is complicated as it involves a sum over many oscillators with different frequencies $(E_n - E_m)$.
Things simplify if we only consider the long-time average, where only terms with $E_n=E_m$ contribute.
\begin{align}
  \bar{g}(\beta) \equiv
  \lim_{t_o \to \infty} \frac{1}{t_o} \int_0^{t_o} g(\beta,t) dt =
  \sum_n N_n^2 e^{-2\beta E_n} \ed
  \label{barg}
\end{align}
We see that, on average, $g(\beta,t)$ approaches a non-zero value at late times.
In \eqref{barg} we implicitly assumed that there is a minimal level spacing in the spectrum.

The long-time average obeys the bound
\begin{align}
  \bar{g}(\beta) \ge Z(2\beta) \ed
  \label{gbound}
\end{align}
The bound is saturated when the spectrum has no degeneracies.\footnote{In 2d CFTs the spectrum always has degeneracies coming from descendant states. It is natural to expect, however, that in a chaotic 2d CFT there are no degenerate primaries.}
In this case the late-time average of $g$, namely $Z(2\beta)$, is exponentially smaller than the initial value $Z^2(\beta)$.
Indeed, in general we have
\begin{align}
  \frac{Z(2\beta)}{Z(\beta)^2} =
  \exp \left[ -2\beta \int_\beta^{2\beta} \frac{S(\beta')}{\beta'^2} d\beta' \right] \ed
\end{align}
For a CFT in $d$ spacetime dimensions the right-hand side is equal to $\exp \left[ - \frac{2}{d} \left( 1 - \frac{1}{2^{d}} \right) S(\beta) \right]$.

\subsection{Information Loss}

We now consider the long-time average $\bar{g}(\beta)$ in the context of the AdS$_3$/CFT$_2$ duality.
Consider a $2d$ CFT on a circle of length $L=2\pi$ that has a holographic bulk dual, and assume as before that the theory has a discrete spectrum.
At high tempereature the thermal state of the theory is dual to a BTZ black hole.
Its partition function is given approximately by $Z(\beta) = \exp \left( 8\pi^2 k / \beta \right)$ where $k=c/24$ and $c$ is the central charge of the field theory.
This is an approximation to the full partition function of the quantum theory.
The BTZ contribution to the spectral form factor can be computed by continuing $\beta \to \beta + it$, and it decays at late times as
\begin{align}
  |Z(\beta,t)|^2
  \sim \exp \left( \frac{16\pi^2 k\beta}{t^2} \right) \ed
\end{align}
In taking the late time limit we will always keep $\beta$ (the real part) fixed.
In this approximation we find that the time average is $1$, violating the bound \eqref{gbound}.
This is a form of information loss.
The BTZ contribution to the partition function is given by the modular image of the vacuum state.
As we will see below, no finite number of additional primary operators is sufficient to avoid information loss.

Let us think clearly about what this means.
Given an approximate partition function $Z(\beta)$ we can compute the corresponding density of states $\rho(E)$ by an inverse Laplace transform.
For the BTZ black hole this is well approximated at high energies by the Cardy formula $\rho_{\rm cardy}(E) = e^{4\pi\sqrt{2kE}}$, which is an approximation to the density of states in the dual field theory.
The important difference between this and the exact density of states \eqref{rhoExact} of the quantum theory is that the Cardy density is a smooth and finite function of the energy (see \cite{Festuccia:2006sa} for a related discussion in the context of large $N$ gauge theories).
Indeed, given a partition function of the form $Z(\beta) = \int dE \rho_s(E) e^{-\beta E}$ where $\rho_s$ is a smooth and finite function, it is easy to see that the time-averaged spectral form factor violates the bound regardless of the details of $\rho_s$.

We see that the late time behavior of the spectral form factor directly probes the discreteness of the spectrum of the theory.
In particular, the time-averaged $\bar{g}(\beta)$ counts discrete states in the spectrum (weighted by a Boltzmann factor and by degeneracy).
Information loss occurs when we approximate the density of states by a smooth function that does not capture the individual energy levels.
This type of information loss occurs in classical black holes in arbitrary dimension.
Equivalently, it occurs in the dual field theory when we use the thermodynamic approximation to the partition function.

\section{$2d$ CFTs}
\label{sec:2d}

In this section we discuss in more detail the torus partition function and spectral form factor in $2d$ CFTs, focusing on theories with large central charge.
We discuss possible corrections to the leading answer (including certain non-perturbative corrections) and show that they are not sufficient to restore information in the spectral form factor.

Consider the partition function of a $2d$ CFT on a torus with parameter $\tau = \frac{i\beta}{2\pi} + \frac{\mu}{2\pi}$.
From now on we set the chemical potential $\mu=0$.
The partition function can be written as a sum over all states,
\begin{align}
  Z(\tau,\bartau) = \sum_{(h,\barh)}
  N_{h,\barh} q^{h-k} \bar{q}^{\barh-k} \ed
\end{align}
Here $q(\tau) \equiv \exp(2\pi i \tau)$, $N_{h,\barh}$ is the degeneracy of the state with conformal weights $(h,\barh)$, and we took the central charges to be $c_L = c_R = c=24 k$ for convenience.
All states have $h,\barh \ge 0$.

The full partition function can also be written as a sum over Virasoro characters,
\begin{align}
  Z(\tau,\bartau) =
  |\chi_0(\tau)|^2 +
  \sum_{(h,\bar{h})} n_{h,\barh} \chi_h(\tau) \bar{\chi}_{\bar{h}}(\bar{\tau})
  \ed
\end{align}
Here we use the notation $\bar{f}(z)=\overline{f(\bar{z})}$.
Each term captures the contribution from a Virasoro primary with dimensions $(h,\bar{h})$ and its descendants, and we have isolated the vacuum contribution from the sum.
Each character appears with degeneracy $n_{h,\barh}$.
The characters are given by
\es{chars}{
  \chi_0(\tau) &= (1-q)\frac{q^{-k+\frac{1}{24}}}{\eta(\tau)} \ec \\
  \chi_h(\tau) &= \frac{q^{h-k+\frac{1}{24}}}{\eta(\tau)} \, \ecq h>0 \ec
}
where $\eta(\tau)$ is the Dedekind eta function.
These expressions are exact even at finite $c$.

We assume the theory is modular invariant, which means
\begin{align}
  Z(\gamma(\tau),\gamma(\bartau)) = Z(\tau,\bartau) \ecq
  \gamma(\tau) = \frac{a\tau + b}{c\tau + d} \ecq
  \gamma \in SL(2,\bZ) \ed
\end{align}
We can write the partition function as a sum over states after performing any $SL(2,\bZ)$ transformation $\gamma$.
We will refer to the $\gamma$-image of a particular character as the contribution of that character in the $\gamma$ frame.

To obtain the high-temperature approximation to the partition function we can write the sum over characters in the $S$ frame.
\begin{align}
  Z(\tau,\bartau) &=
  \chi_0(-1/\tau)\bar{\chi}_0(-1/\bar{\tau}) +
  \sum_{(h,\bar{h})} n_{h,\barh} \chi_h(-1/\tau)
  \bar{\chi}_{\bar{h}}(-1/\bar{\tau}) 
  \ed \label{ZSframe}
\end{align}
The first term, which is the vacuum character contribution in the $S$ frame, is the dominant contribution at high temperatures (when $\beta<2\pi$) \cite{Hartman:2014oaa}.
It is given by
\begin{align}
  Z_{\rm BTZ}(\tau,\bar{\tau}) \equiv
  \chi_{0}(-1/\tau)\bar{\chi}_{0}(-1/\bar{\tau}) =
  \frac{2\pi}{\beta} \frac{1}{|\eta(\tau)|^2}
  \left( 1 - e^{-4\pi^2/\beta} \right)^2
  \exp \left[
    \frac{8\pi^2}{\beta} \left(k-\frac{1}{24} \right)
  \right] \ed \label{ZBTZ}
\end{align}
In writing this we used the fact that $\eta(-1/\tau) = \sqrt{-i\tau} \cdot \eta(\tau)$.
The leading part of \eqref{ZBTZ} at large $c$ comes from the vacuum state itself.
It also has an $\cO(1/c)$ correction coming from the sum over the vacuum's descendants.

Much of this structure is echoed on the gravitational side.
The asymptotic symmetry algebra of pure gravity in AdS$_{3}$ is the Virasoro algebra, with central charge $c=\frac{3 \ell}{2 G}$ \cite{Henneaux:1985tv}.\footnote{Here $\ell$ is the AdS length, and $G$ is the $3d$ Newton constant.}
The contribution of thermal AdS$_3$ to the partition function can be evaluated exactly, and is given by the vacuum character contribution
\es{Zvac}{
  Z_{\rm vac}(\tau,\bar{\tau}) &\equiv
  \chi_{0}(\tau)\bar{\chi}_{0}(\bar{\tau})\,.
}
In the bulk, the leading contribution comes from evaluating the action of the classical gravity solution, while the $\cO(1/c)$ correction is due to a 1-loop determinant.
There are no higher order corrections so this result is 1-loop exact in bulk
language.

The contribution of the BTZ black hole geometry is given by the vacuum character in the $S$ frame, eq.
\eqref{ZBTZ}.
Here, again, the leading large $c$ contribution comes from the classical (black hole) solution, and there is an $\cO(1/c)$ 1-loop determinant.

More generally, as we review in Appendix~\ref{app:bh}, at fixed temperature and chemical potential there are an infinite number of classical bulk solutions that are related by $SL(2,\mathbb{Z})$ transformations \cite{Maldacena:1998bw}.\footnote{The family of solutions is labeled by elements of $\Gamma_{\infty}\backslash SL(2;\mathbb{Z})$, where we quotient by $\tau\rightarrow\tau+1$ on the left.}
The solution that corresponds to $\gamma \in SL(2,\mathbb{Z})$ makes a contribution to the partition function equal to $\chi_{0}(\gamma(\tau))\bar{\chi}_{0}(\gamma(\bar{\tau}))$.
For a general theory, there will be many additional contributions to the partition function that correspond to states involving matter fields.

\subsection{Analytic Continuation to Real Time}

Equation \eqref{ZSframe} is a useful starting point for the analytic continuation $\beta \to \beta + it$ to real time because (i) at $t=0$ the vacuum character contribution provides a good approximation, and (ii) this dominant contribution has a clear bulk interpretation as the BTZ black hole.
This contribution remains dominant at sufficiently early times.
We now discuss the various pieces of eq.
\eqref{ZSframe} after analytic continuation to late times.
We will find that the contribution coming from each individual character decays to zero at late times, violating the bound \eqref{gbound}.

We start by focusing on the contribution of the vacuum character \eqref{ZBTZ}, which is equal to the BTZ black hole contribution and is the dominant contribution at high temperature.
We analytically continue $\beta$, taking
\begin{align}
  \tau = \frac{i(\beta+it)}{2\pi} \ecq
  \bar{\tau} = -\frac{i(\beta+it)}{2\pi} \ed
  \label{prams0}
\end{align}
Notice that after analytic continuation $\bar{\tau}$ is not the complex conjugate of $\tau$.
After a time of order a few $\beta$s we find that the vacuum character contribution to the spectral form factor decays as\footnote{
  The continued eta function $\eta(\tau)$ oscillates in time, never giving a substantial contribution to \eqref{gbtz}.
  }
\es{gbtz}{
  |Z_{\rm BTZ}(\tau,\bar{\tau})|^2 
  &\sim \frac{1}{t^6}
  \exp \left( \frac{16 \pi ^2 \beta}{\beta ^2+t^2} k
  \right)
  \,.
}
The leading, vacuum state contribution decays exponentially to an $\cO(1)$ amplitude at times $t~\sim~\sqrt{k}$.
The subleading contribution coming from the descendents is then responsible for the $1/t^6$ power law decay down to zero.
Curiously, including additional states (the vacuum's descendents) in the $S$-frame makes the violation of the bound \eqref{gbound} \textit{worse}.

Next, the contribution to \eqref{ZSframe} coming from each non-vacuum character can be written as
\begin{align}
  \chi_h(-1/\tau) \bar{\chi}_{\bar{h}}(-1/\bar{\tau}) =
  \frac{2\pi}{\beta} \frac{1}{|\eta(\tau)|^2}
  \exp \left[
    \frac{8\pi^2}{\beta}
    \left(k-\frac{1}{24} - \frac{h + \bar{h}}{2}\right)
  \right]
  \ed \label{chihS}
\end{align}
(In writing this we assumed for simplicity that neither $h$ nor $\bar{h}$ are equal to zero, \textit{i.e.} we are excluding additional conserved currents.)

Regardless of the conformal dimensions, the character decays in time as $1/t$, with a pre-factor which decreases as $h,\bar{h}$ increase.
We arrive at the following conclusion: Including a finite number of characters in the $S$-frame does not bring us closer to obeying the bound $\eqref{gbound}$.

\section{Universal Late Time Decay}
\label{sec:uni}

In this section we will attempt to understand universal properties of the late time partition function in AdS$_{3}$/CFT$_{2}$.
For gravity in weakly curved AdS$_{3}$ the partition function undergoes a phase transition between the dominant low temperature saddle, thermal AdS$_3$, and the high temperature saddle, the BTZ black hole \cite{Hawking:1982dh}.
The partition function in these two regimes is given approximately by
\es{Zeucphase}{
\log Z(\beta)&=\left\{\begin{array}{ll}2\beta k&,\,\beta>2\pi\\ \frac{8\pi ^2 k}{\beta }&,\,\beta<2\pi \end{array}\right. +\mathcal{O}(1)\,.
}
This phase structure is replicated in sufficiently sparse, large $c$ CFTs \cite{Hartman:2014oaa}.
As long as the number of states grows sub-exponentially, the partition function is dominated by the vacuum state at low temperatures, and by the vacuum state in the $S$ frame at high temperatures.
As discussed above, starting with the dominant high temperature contribution and continuing $\beta$ analytically to real time does not reproduce the correct late time behavior.
The spectral form factor satisfies the bound \eqref{gbound},
\es{gunit}{
  \bar{g}(\beta)=\lim_{t_o \to \infty} \frac{1}{t_o} \int_0^{t_o} |Z(\beta+it)|^{2} dt \geq Z(2\beta) \ec
}
while the thermal partition function corresponding to the BTZ black hole leads to a decaying spectral form factor \eqref{gbtz}.
This contribution decays exponentially to an $\cO(1)$ amplitude at times of order $\sqrt{k}$.
As we will show, this decay significantly underestimates the correct late time behavior of the partition function.

In Section~\ref{sec:UC} we identify a universal contribution to the partition function which decays significantly slower than \eqref{gbtz}.
Then, in Section~\ref{sec:dom} we estimate corrections to the universal decay using Cardy's formula, and find that they are negligible in this approximation.
In Section~\ref{sec:rathot} we give a refined version of the universal contribution to the partition function for all times and temperatures.
This, together with the late time plateau for the spectral form factor, lends evidence to a universal picture for the time dependence of the partition function that we lay out in Section~\ref{sec:dip}.

\subsection{Universal Contribution}\label{sec:UC}

The partition function \eqref{ZSframe} expanded in the $S$ frame is dominated by the vacuum character at $t=0$.
This suggests a strategy for approximating the partition function at later times: At any given time, identify the apropriate modular transformation such that the image of the vacuum character in this frame is larger than in any other frame.

Consider the partition function at times $t_{n}\equiv 2\pi n$, with corresponding modular parameters
\es{prams}{
  \tau_{n}\,=\,\frac{i(\beta+it_{n})}{2\pi}\,, \quad
  \bar{\tau}_{n}\,=\,-\frac{i(\beta+it_{n})}{2\pi}\,, \quad
  n\in\mathbb{Z}\,.
}
To study the partition function at these discrete times, it is convenient to perform a time-dependent modular transformation $\gamma_{n}(\tau)\equiv-1/(\tau+n)$.
\begin{align}
  \gamma_n(\tau_n) = \frac{2\pi i}{\beta} \ecq
  \gamma_n(\bar{\tau}_n) = - \frac{2\pi i}{\beta + 2it_n} \ed
\end{align}
This transformation removes all of the holomorphic time dependence.
It maximizes the contribution from the vacuum character among all modular transformations.
Explicitly, the vacuum character in the $\gamma_n$ frame is given by 
\begin{align}
  Z_{\rm vac}(\gamma_{n}(\tau_{n}),\gamma_{n}(\bar{\tau}_{n})) &=
  \chi_{0} \left(\frac{2 \pi i}{\beta} \right)
  \bar{\chi}_{0}
  \left( -\frac{2 \pi i}{\beta +4 i \pi n} \right)
  \cr
  &=
  \sqrt{\frac{4 \pi^2 }{\beta(\beta -4 \pi  in)}} \cdot
  \frac{
    \exp \left[ 4 \pi^2 \left( k - \frac{1}{24} \right) \left(
      \frac{1}{\beta} +
      \frac{1}{\beta + 4 \pi i n}
    \right) \right]
  }{
    \left| \eta \left(\frac{i\beta }{2\pi }\right) \right|^2
  }
  \left(1-e^{-\frac{4 \pi ^2}{\beta }}\right)
  \left(1-e^{-\frac{4 \pi ^2}{\beta +4 \pi i n}}\right)
  \,.
  \cr
\end{align}
It decays at late times (large $n$) as
\begin{align}
  Z_{\rm vac}(\gamma_{n}(\tau_{n}),\gamma_{n}(\bar{\tau}_{n})) &\sim
  \frac{e^{4 \pi ^2 k / \beta}}{t_n^{3/2}} \,. \label{Zn}
\end{align}
Notice that the vacuum state itself decays in this frame to the exponentially large value $e^{4 \pi ^2 k / \beta}$, which is much larger than the asymptotic value of the vacuum state in the $S$ frame.
The power law decay is due entirely to the $\cO(1/c)$ piece of the vacuum character.

\begin{figure}
  \centering
    \includegraphics[width=0.8\textwidth]{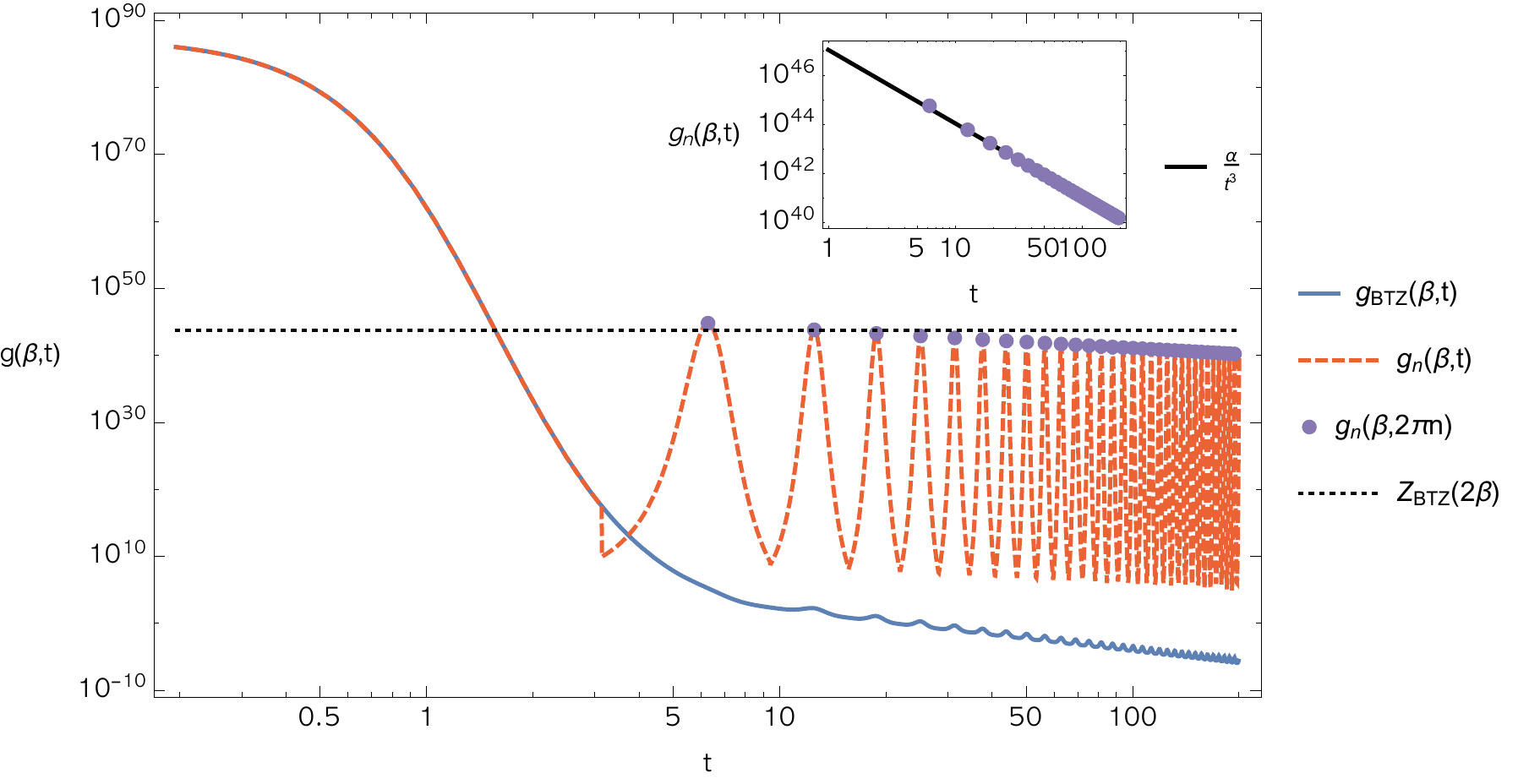}
    \caption{The spectral form factors corresponding to the BTZ black hole contribution $g_{\rm BTZ}(\beta,t)$ (blue) and corresponding to the dominant image of the vacuum $g_{n}(\beta,t)$ (red). Here, for $t\neq2\pi n$, we interpolate by taking $n={\rm integer\, part}(t/2\pi)$. This accounts for the discontinuities in the red, dashed line.
The peaks of this contribution are attained at discrete times $t_n$ (purple dots).
Going to the late dominant frame does not avoid late time decay, violating the late time bound \eqref{gbound} (black, dotted).
Inset: The dominant contribution at $t_n$ (purple) with a fit to a $t^{-3}$ power law (black).
    }
\label{fig:guniv}
\end{figure}

The vacuum character contribution to the spectral form factor in this frame then decays as
\begin{align}
  g_n(\beta,t_n) \equiv
  |Z_{\rm vac}(\gamma_{n}(\tau_{n}),\gamma_{n}(\bar{\tau}_{n}))|^2 &\sim
  \frac{e^{8 \pi ^2 k / \beta}}{t_n^{3}} \label{gn} \,.
\end{align}
Figure~\ref{fig:guniv} shows $g_n(\beta,t)$ compared with the late time bound \eqref{gbound} and the decay from the vacuum character in the $S$ frame.
Notice that the amplitude of the power law decay in \eqref{gn} is in fact greater than the value of the late time bound \eqref{gbound}, $Z(2\beta) \approx \exp \left( \frac{4\pi^2 k}{\beta} \right)$.
Next, let us consider the contribution of a non-vacuum character $Z_{h,\bar{h}}(\tau,\bar{\tau}) = \chi_h(\tau) \bar{\chi}_{\bar{h}}(\bar{\tau})$.
We assume that the state is `light', namely that the conformal weights $h,\bar{h}$ are fixed as we take $k$ large.
We will also assume for simplicity that there are no extra currents, \textit{i.e.} $h,\bar{h}$ are both strictly positive.
At time $t_n$ the $\gamma_n$ frame again maximizes the contribution of the character among all $SL(2,\bZ)$ frames.
At late times this contribution to the partition function decays as
\begin{align}\label{domrel}
  Z_{h,\bar{h}}(\gamma_n(\tau_n),\gamma_n(\bar{\tau}_n)) \sim
  \frac{e^{4\pi^2 k/\beta}}{t_n^{1/2}} \ed
\end{align}
The faster decay compared with the vacuum character \eqref{Zn} can be traced back to the fact that the vacuum character has an additional $(1-q)$ factor that decays as $1/t_n$.
The matter character contribution to the spectral form factor behaves at late times as
\begin{align}
  g_n^{(h,\bar{h})}(\beta,t_n) \equiv
  |Z_{h,\bar{h}}(\gamma_n(\tau_n),\gamma_n(\bar{\tau}_n))|^2 \sim
  \frac{e^{8\pi^2 k/\beta}}{t_n} \ed
  \label{gnmat}
\end{align}

In Section~\ref{sec:rathot} we will generalize these considerations and find a universal contribution to the spectral form factor for arbitrary rational times.
The result will be bounded above by \eqref{gnmat} if we replace $t_n$ by a rational time.
We conjecture that the universal contributions from the vacuum \eqref{gn} together with the contribution from the light states \eqref{gnmat} correctly describe the spectral form factor for generic chaotic CFTs up to the dip time.
For a putative CFT that is dual to pure gravity there are no light matter fields, and we conjecture that correct description is given by \eqref{gn}.
We provide an argument for this in the next subsection.
As discussed in the introduction, beyond the dip time we expect another universal contribution, one due to random matrix theory, to become dominant and lead to a ramp and a plateau.

This universal contribution we have identified, \eqref{gnmat}, has a nice connection with classical bulk saddles. As we review in Appendix~\ref{app:bh}, for each $n$ there is a black hole solution in the bulk, with the contribution $Z_{\rm vac}(\gamma_{n}(\tau),\gamma_{n}(\bar{\tau}))$ to the gravitational partition function. We can thus identify the universal decay of the spectral form factor with the contribution of these black hole solutions.

\subsection{Dominance of the Universal Contribution}
\label{sec:dom}

In the previous subsection we identified a universal contribution to the partition function.
We now argue that this contribution provides a good approximation to the partition function before the dip time, namely before the universal contribution due to random matrix theory becomes dominant.

Focusing again on the discrete times $t_{n}=2\pi n$ the full partition function can be written as a sum over states in the $\gamma_n$ frame,
\es{dominanceeq}{
Z(\beta+i t_{n})\,=\,Z(\gamma_{n}(\tau_{n}),\gamma_{n}(\bar{\tau}_{n}))\,=\,e^{\frac{4 \pi ^2 k}{\beta }}e^{\frac{4\pi ^2 k}{\beta+4 \pi  i n}}\left(1+\sum_{h,\bar{h}>0}N_{h,\bar{h}}e^{\frac{-4 \pi ^2 h}{\beta }}e^{\frac{-4\pi ^2 \bar{h}}{\beta+4 \pi  i n}}\right)\,.
}
The factor in front on the right-hand side is equal to the vacuum state contribution in the $\gamma_n$ frame.
This is the amplitude of the universal contribution \eqref{Zn}.
Our goal is to argue that the sum \eqref{dominanceeq} is well approximated by the universal contribution, \eqref{Zn}, until the dip time. We begin by explaining why the sum over the heavy states gives a subdominant contribution to the partition function, and then why the light states and descendants reproduce the amplitude and power law decay of \eqref{Zn}.

The correction to the leading amplitude in \eqref{dominanceeq} is
\begin{align} 
f&\equiv\sum_{h,\bar{h}>0}N_{h,\bar{h}}e^{\frac{-4 \pi ^2 h}{\beta }}e^{\frac{-4\pi ^2 \bar{h}}{\beta+4 \pi  i n}}
\cr
&=
\underbrace{\sum_{h<k||\bar{h}<k}N_{h,\bar{h}}e^{\frac{-4 \pi ^2 h}{\beta }}e^{\frac{-4\pi ^2 \bar{h}}{\beta+4 \pi  i n}}}_{f_{L}}+
\underbrace{\sum_{h,\bar{h}>k}N_{h,\bar{h}}e^{\frac{-4 \pi ^2 h}{\beta }}e^{\frac{-4\pi ^2 \bar{h}}{\beta+4 \pi  i n}}}_{f_{H}}
\label{boundfac}
\end{align}
In the second line we separated the sum over all states into a sum $f_L$ over `light' states, and a sum $f_H$ over `heavy' states.
Let us discuss these two sums separately.

\paragraph{Heavy states.} 
We consider first the sum over heavy states, which we can write as
\begin{align}
  f_H =
  e^{\frac{-4 \pi ^2 k}{\beta }}e^{\frac{-4\pi ^2 k}{\beta+4 \pi  i n}}
  \int_0^\infty \! d\hat{h} \, d\hat{\bar{h}} \,
  \rho(\hat{h}, \hat{\bar{h}})
  e^{\frac{-4 \pi ^2 \hat{h}}{\beta }}
  e^{\frac{-4\pi ^2 \hat{\bar{h}}}{\beta+4 \pi  i n}}
  \ed \label{fH}
\end{align}
Here $\hat{h}\equiv h-k$, and $\rho(\hat{h}, \hat{\bar{h}})$ is the density of heavy states.
This density of states can be approximated by the Cardy density $\rho_c$ \cite{Cardy:1986ie}, which is defined by the equation
\begin{align}\label{cardy}
  e^{2\pi i k\left(\frac{1}{\tau}-\frac{1}{\bar{\tau}}\right)}&=
  \int_{0}^{\infty} \! d\hat{h} \, d\hat{\bar{h}} \,
  \rho_c(\hat{h},\hat{\bar{h}}) e^{2\pi i \tau\hat{h}}e^{-2\pi i \bar{\tau}\hat{\bar{h}}} \ed
\end{align}
The integral on the right is exactly the integral that appears on the right-hand side of \eqref{fH} if we approximate the full density of states $\rho$ by the Cardy density $\rho_c$, and replace $\tau=\frac{4\pi^{2}i}{\beta}$ and $\bar{\tau}=-\frac{4\pi^{2}i}{\beta+2i t_{n}}$.
Therefore, in the Cardy approximation we find that
\begin{align}
  f_H \approx
  e^{\frac{-4 \pi ^2 k}{\beta }}e^{\frac{-4\pi ^2 k}{\beta+2 i t_n}}
  e^{\frac{k \left(\beta + i t_n\right)}{\pi}} \ed
\end{align}
In the large $k$, high temperature limit we see that $f_H \ll 1$ at arbitrarily late times, and so the contribution from the heavy states cannot significantly change the amplitude in \eqref{dominanceeq}.

It is instructive to verify that this suppression of heavy states does not rely on detailed properties of the Cardy distribution.
The solution to \eqref{cardy} is
\begin{align}
  \rho_c(\hat{h},\hat{\bar{h}}) &=
  \rho_c(\hat{h}) \rho_c(\hat{\bar{h}}) \ec \\
  \rho_c(\hat{h}) &= 
  \delta(\hat{h}) + 2\pi \sqrt{\frac{k}{\hat{h}}} \cdot
  I_1\left( 4\pi \sqrt{k\hat{h}} \right) \cr &=
  \delta(\hat{h}) +
  \left( \frac{k}{4 \hat{h}^3} \right)^{1/4}
  e^{4\pi \sqrt{k\hat{h}}} \left[ 1 + \cO(\hat{h}^{-1/2}) \right] \ed
\end{align}
In the last line we expanded to leading order in large $\hat{h}$.
It is easy to check that this leading piece (including the $\hat{h}^{-3/4}$ factor) also leads to a suppressed contribution from the heavy states.

\paragraph{Light states.}
The contribution from light states is more subtle. To constrain the contribution of the light states, we would like to appeal to sparsity.
In other words, we would like to consider theories without too many light states. However, we always have, at the very least, Virasoro descendants of the vacuum. As the light state contribution,
\begin{align}
  f_L =
  \int_{h<k||\bar{h}<k} \! d{h} \, d{\bar{h}} \,
  \rho({h}, {\bar{h}})
  e^{\frac{-4 \pi ^2 {h}}{\beta }}
  e^{\frac{-4\pi ^2 {\bar{h}}}{\beta+4 \pi  i n}}
  \ec \label{fL}
\end{align}
has no suppression, it is difficult to argue that the light states give an $\mathcal{O}(1)$ contribution at late times. Indeed, if this were the case, it would contradict the power law decay of our universal contribution \eqref{gnmat}. To address this fact, and to give teeth to the assumption of sparsity, we turn our attention to the expansion of the partition function in terms of characters rather than states.

\paragraph{Light and heavy characters.}
The universal contribution \eqref{gnmat} contains an amplitude and a subleading power-law decay, which comes from summing over descendants.
The descendants include heavy states which contribute to the Cardy relation \eqref{cardy}.
To show the dominance of the full contribution \eqref{gnmat} (including the power law decay) we re-expand the partition function in characters instead of in states, in the $\gamma_n$ frame.
We define $\sigma_n \equiv \gamma_n(\tau_n)$ and $\bar{\sigma}_n \equiv \gamma_n(\bar{\tau}_n)$ to reduce clutter.
\begin{align}
  Z(\beta + it) = Z(\sigma_n, \bar{\sigma}_n) &=
  \chi_0(\sigma_n) \bar{\chi}_0(\bar{\sigma}_n)) 
  + \cr &\quad
  \sum_{h<k||\bar{h}<k} n_{h,\bar{h}}
  \chi_h(\sigma_n) \bar{\chi}_{\bar{h}}(\bar{\sigma}_n)
  + \cr &\quad
  e^{\frac{-4 \pi ^2 k}{\beta }}e^{\frac{-4\pi ^2 k}{\beta+4 \pi  i n}}
  \int_0^\infty \! d\hat{h} \, d\hat{\bar{h}} \,
  \rho_{\chi}(\hat{h}, \hat{\bar{h}})
  \chi_{\hat{h}}(\sigma_n)
  \bar{\chi}_{\hat{\bar{h}}}(\bar{\sigma}_n) \ed
  \label{Zsig}
\end{align}
Here $\rho_\chi(\hat{h}, \hat{\bar{h}})$ denotes the density of characters with conformal dimensions $(h,\bar{h})$, and we took out factors of $q^k$ as in \eqref{fH}.
As before, $\hat{h} \equiv h - k$.
The term $\chi_0(\sigma_n) \bar{\chi}_0(\bar{\sigma}_n))$ is the universal vacuum contribution \eqref{gn}.
The sum on the second line is the contribution from light characters.

The primaries we are describing as light here consist of any state with either $h$ or $\bar{h}$ smaller than $k$.
These are referred to as censored primaries in \cite{Keller:2014xba}.
One way to justify limiting the number of such states, is that those with either $h\gg \bar{h}$ or $\bar{h}\gg h$ are close to conserved currents, and we expect there to be few such states in a typical chaotic CFT.
More generally, we would like to consider CFTs that are dual to gravitational theories without too much matter.
For us, sparseness means simply that the the contribution from these light primaries is well approximated by the vacuum character, with at most an order one number of additional light primaries.\footnote{Note, this is more strict then what is sometimes imposed (see \cite{Hartman:2014oaa} for instance), and requires a separation of scales between the AdS length and the string scale in the bulk.}

Finally, on the last line we have the contribution of the heavy characters, which we claim is negligible in the Cardy approximation.
We can approximate the density of the heavy characters by a Cardy density $\rho_\chi \approx \rho_{\chi,c}$, which is defined by the equation
\begin{align}\label{cardyChi}
  e^{2\pi i k\left(\frac{1}{\tau}-\frac{1}{\bar{\tau}}\right)}&=
  \int_{0}^{\infty} \! d\hat{h} \, d\hat{\bar{h}} \,
  \rho_{\chi,c}(\hat{h},\hat{\bar{h}})
  \chi_{\hat{h}}(\tau)
  \bar{\chi}_{\hat{\bar{h}}}(\bar{\tau}) \ed
\end{align}
As in the case of heavy states, the integral on the right-hand side is the same integral that appears in \eqref{Zsig}, and the same argument implies that this contribution will be negligible.

The arguments above seem to imply a decaying spectral form factor at arbitrarily late times, but we know that they must fail at some point in order for the lower bound \eqref{gbound} on the plateau height to be satisfied.
In particular, the assumption that the density of characters is well approximated by the Cardy density becomes invalid at sufficiently late times.
The left-hand side of \eqref{cardyChi} includes only the vacuum state.
In the full theory the left-hand side includes other states, whose contribution becomes important at late times.

In this work we assume that at late times the only important physical effects are the universal decay before the dip time, and the random matrix theory behavior of a ramp + plateau beyond it.
This is equivalent to assuming that the density of characters $\rho_\chi$ is well approximated by the Cardy density until the dip time.

\subsection{Rational Times and Hot Saddles}
\label{sec:rathot}
So far we focused on the discrete times $t_{n}=2\pi n$.
The story at generic times is slightly more elaborate.
We begin by considering the times $t_{n+1/2}=2\pi(n+1/2)$, $n \in \bZ$, and the corresponding modular parameters $\tau_{n+1/2}$ and $\bar{\tau}_{n+1/2}$.
There are now two modular transformations of the vacuum that vie for dominance at high temperatures: $\gamma_n$ and $\gamma_{2,2n+1}$, where we define $\gamma_{c,d}(\tau)\equiv \frac{a \tau+b}{c\tau+d}$ (where $a,b$ are uniquely determined from $c,d$).
Indeed, we have our previous choice,
\es{znhalftime}{
  Z_{\rm vac}(\gamma_{n}(\tau_{n+1/2}),
  \gamma_{n}(\bar{\tau}_{n+1/2})) &=
  \exp \left[ 
  \frac{4 \pi ^2 k}{\beta +i \pi } +
  \frac{4 \pi ^2 k}{\beta -4 i \pi  n-i \pi }
  + \cO(k^0)
\right] \,.
}
And we have the competing modular frame,
\es{znhalftime}{
  Z_{\rm vac}(\gamma_{2,2n+1}(\tau_{n+1/2}),
  \gamma_{2,2n+1}(\bar{\tau}_{n+1/2})) &=
  \exp \left[
  \frac{\pi  (\pi -i \beta ) k}{\beta }
  -\frac{i \pi  k (-i \beta +4 \pi  n+\pi )}{\pi  (4 n+2)-i \beta }
  + \cO(k^0)
\right] \,.
}
At late times (large $n$) we compare the two contributions,
\begin{gather}
  \log\left|Z_{\rm vac}(\gamma_{n}(\tau_{n+1/2}),\gamma_{n}(\bar{\tau}_{n+1/2}))\right| \approx\frac{4 \pi ^2 \beta  k}{\beta ^2+\pi ^2} + \cO(k^0) \,,~\mathrm{and}\\
  \log\left|Z_{\rm vac}(\gamma_{2,2n+1}(\tau_{n+1/2}),\gamma_{2,2n+1}(\bar{\tau}_{n+1/2}))\right| \approx\frac{\pi ^2 k}{\beta } + \cO(k^0) \,.
  \label{exchangesad}
\end{gather}
For sufficiently high temperature, $\beta<\frac{\pi}{\sqrt{3}}$, the second contribution is larger and gives the dominant contribution, while for $\frac{\pi}{\sqrt{3}}<\beta<2\pi$ the first contribution dominates.

More generally, for any rational time, $t_{n/m}=\frac{2\pi n}{m}$, there exists an inverse temperature, $\beta_{m,n}$, such that for $\beta<\beta_{m,n}$, the vacuum in the modular frame $\gamma_{m,n}$ gives a bigger contribution than the vacuum in any other frame.

\begin{figure}
  \centering
    \includegraphics[width=0.8\textwidth]{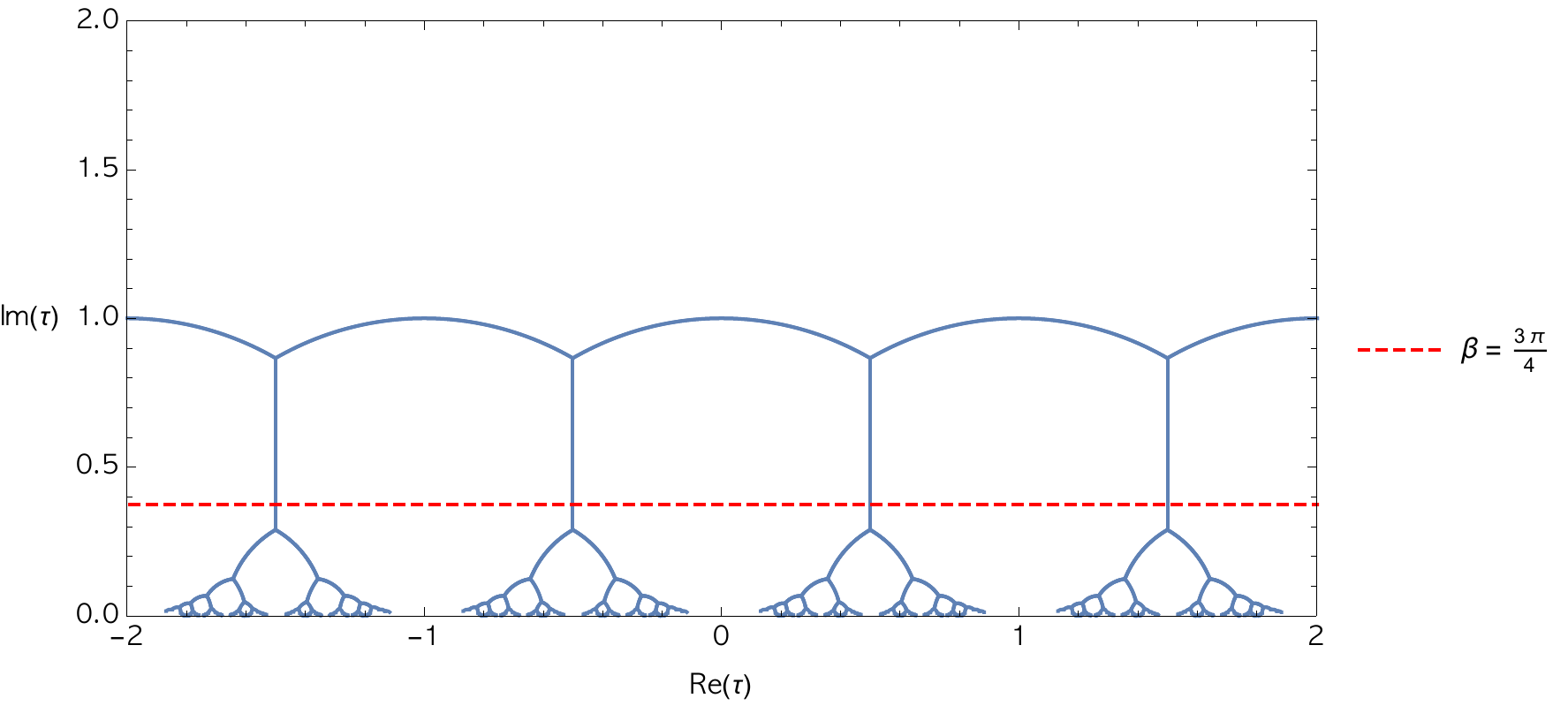}
    \caption{Here, we show the upper half-plane tiled by fundamental domains of $SL(2;\mathbb{Z})$. As we increase the temperature, which corresponds to lowering the red line, we cross more and more fundamental domains.}
\label{fig:funddom}
\end{figure}

\begin{figure}
  \centering
  	\includegraphics[width=0.24\textwidth]{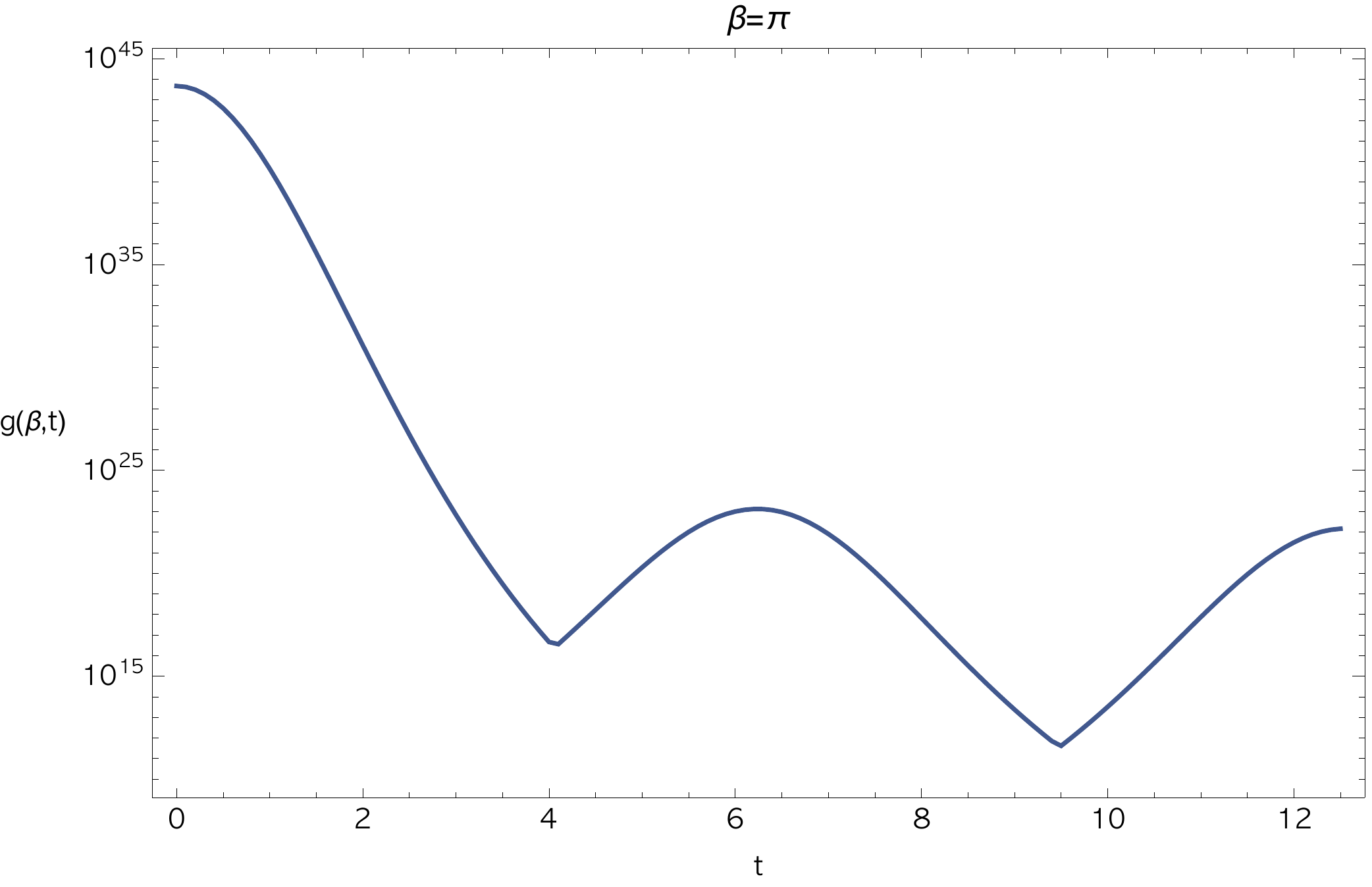}
	\includegraphics[width=0.24\textwidth]{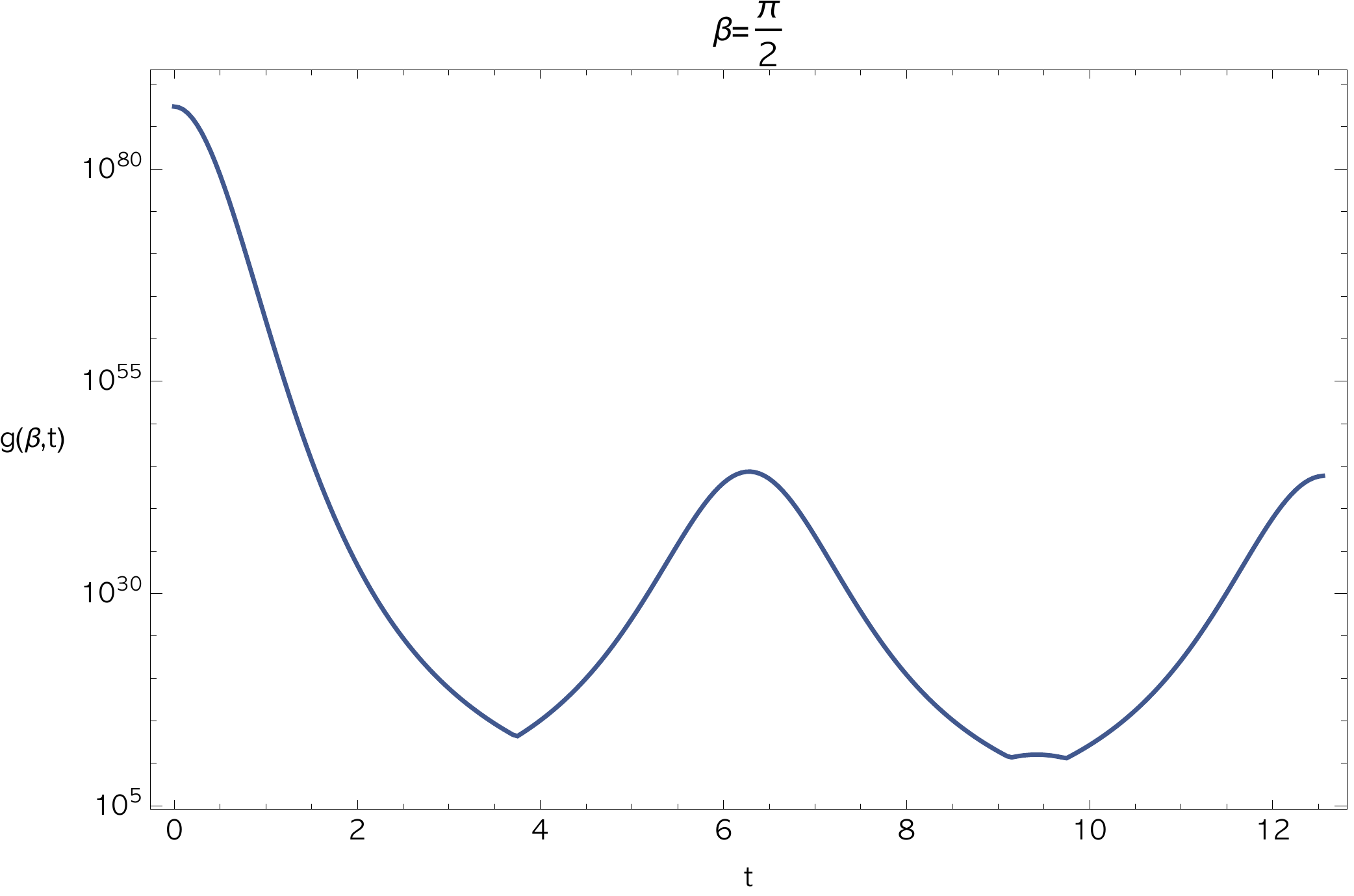}
	\includegraphics[width=0.24\textwidth]{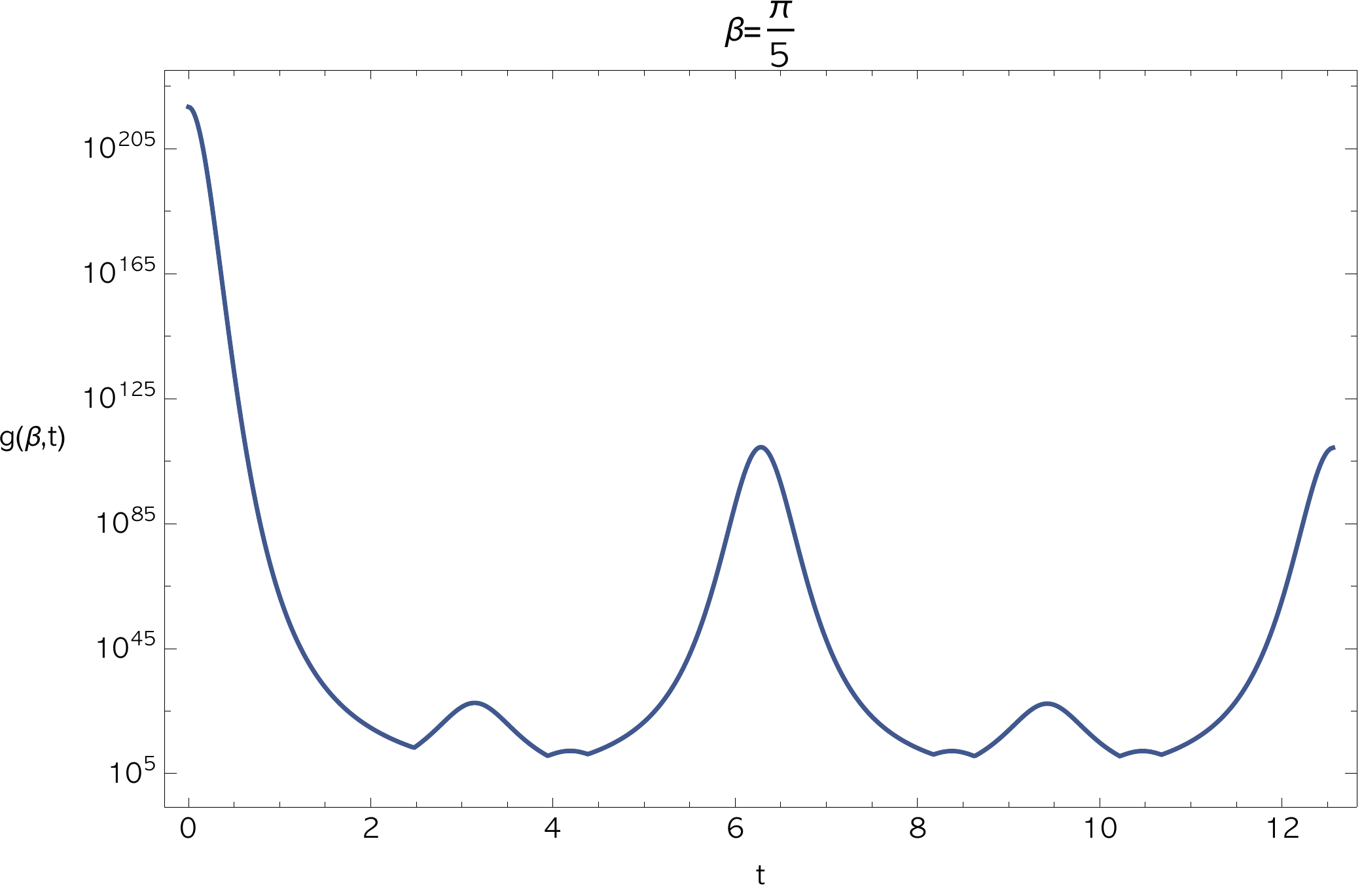}
	\includegraphics[width=0.24\textwidth]{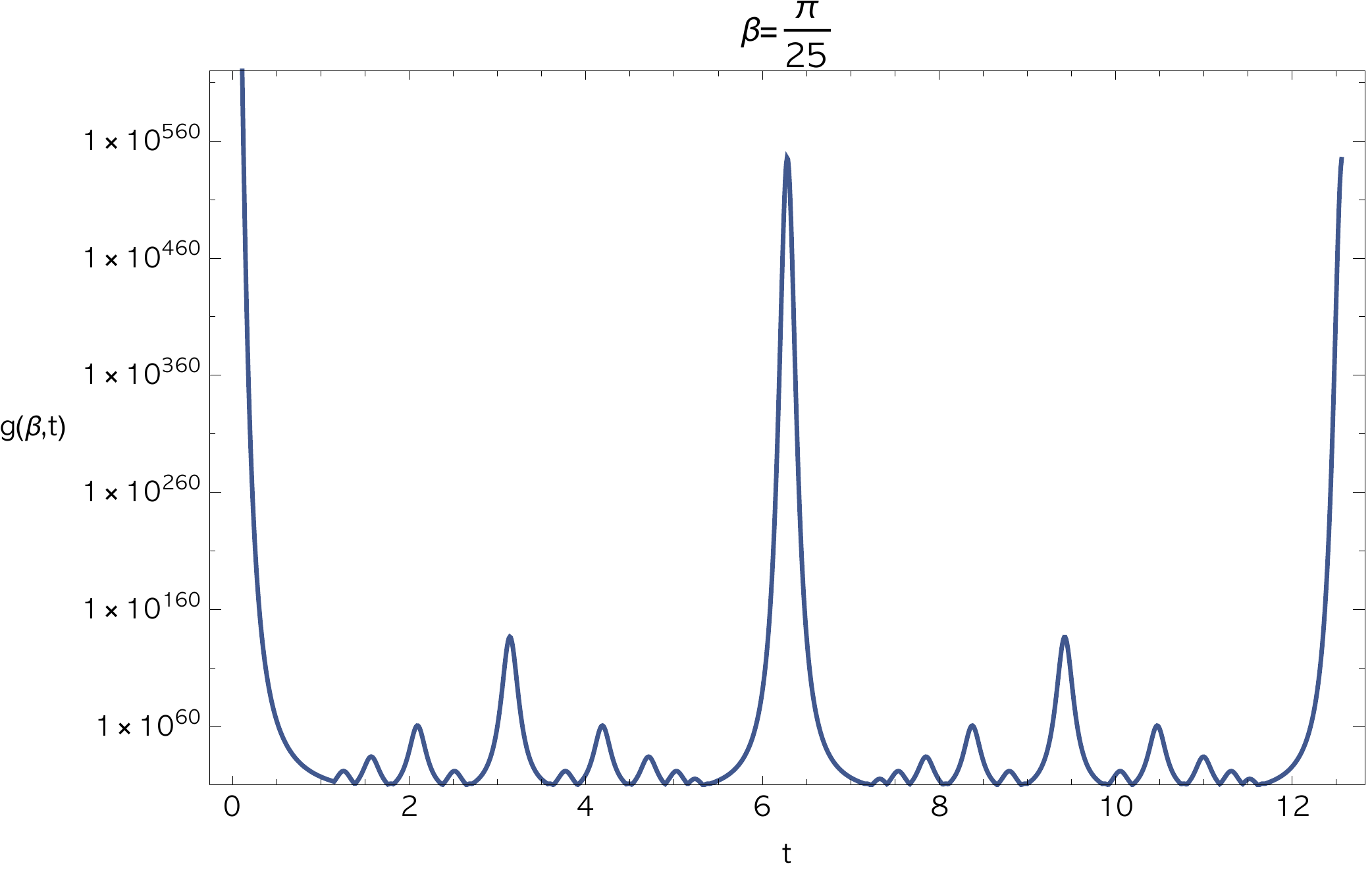}\\
	\includegraphics[width=0.24\textwidth]{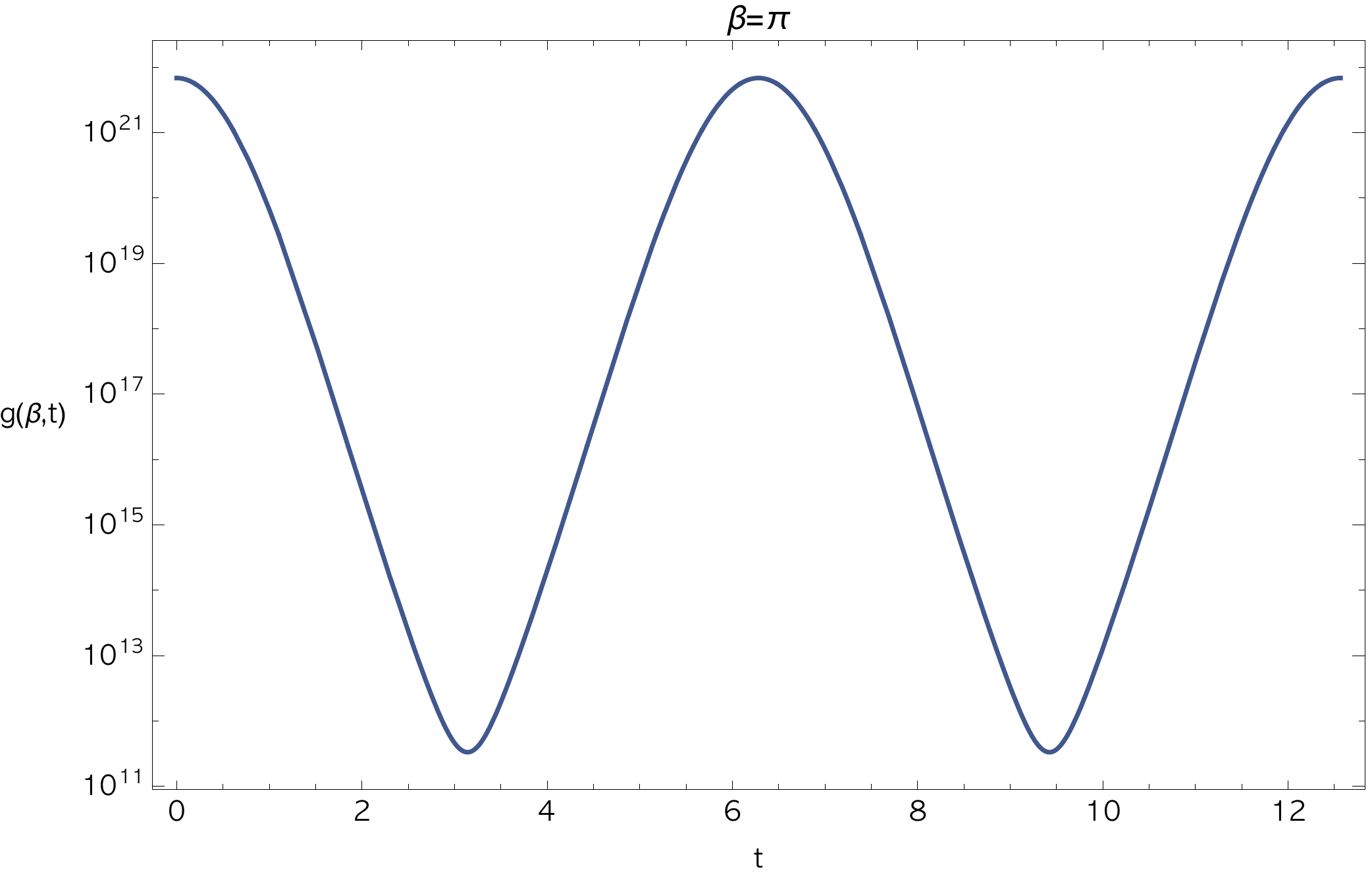}
	\includegraphics[width=0.24\textwidth]{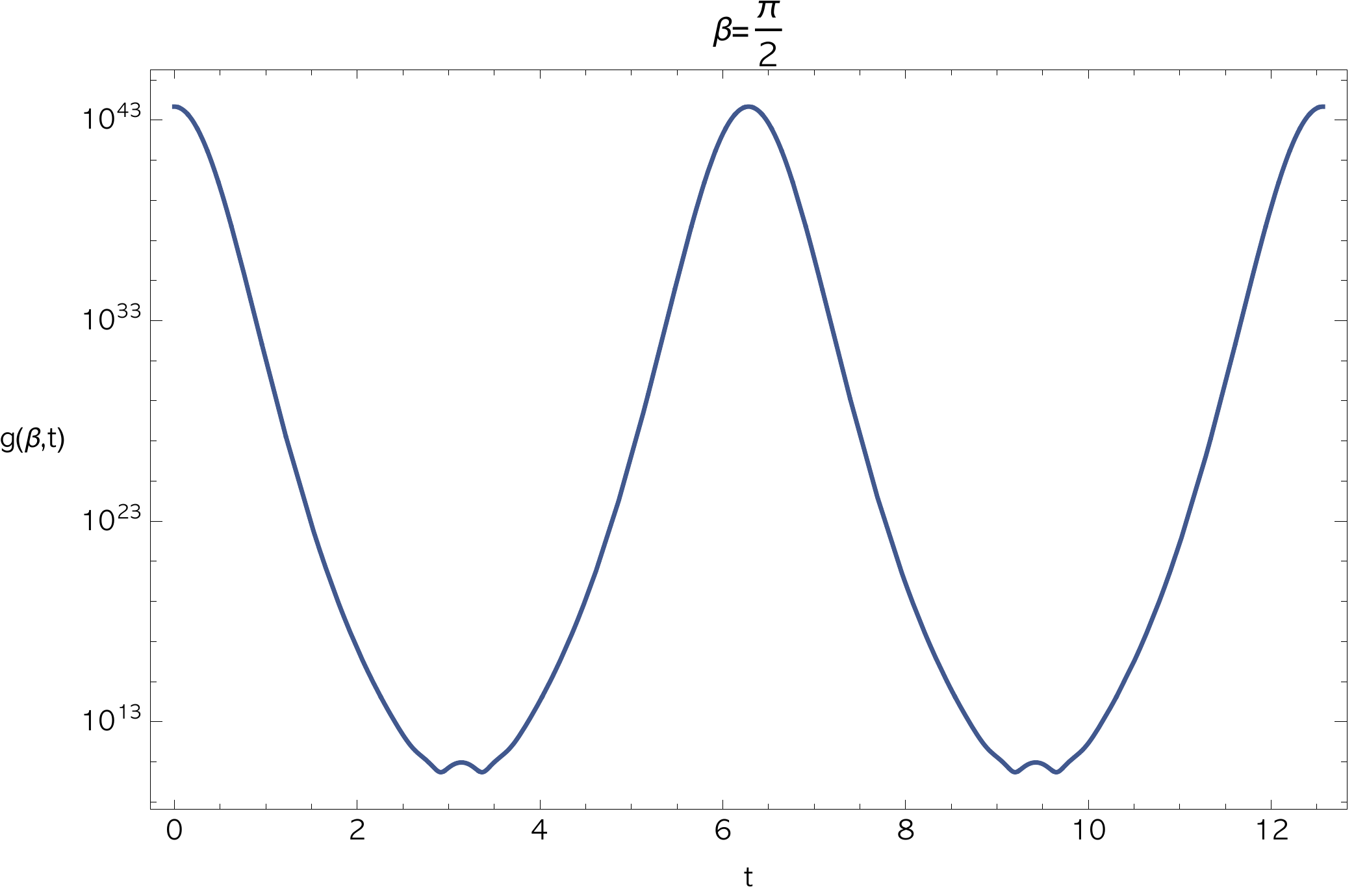}
	\includegraphics[width=0.24\textwidth]{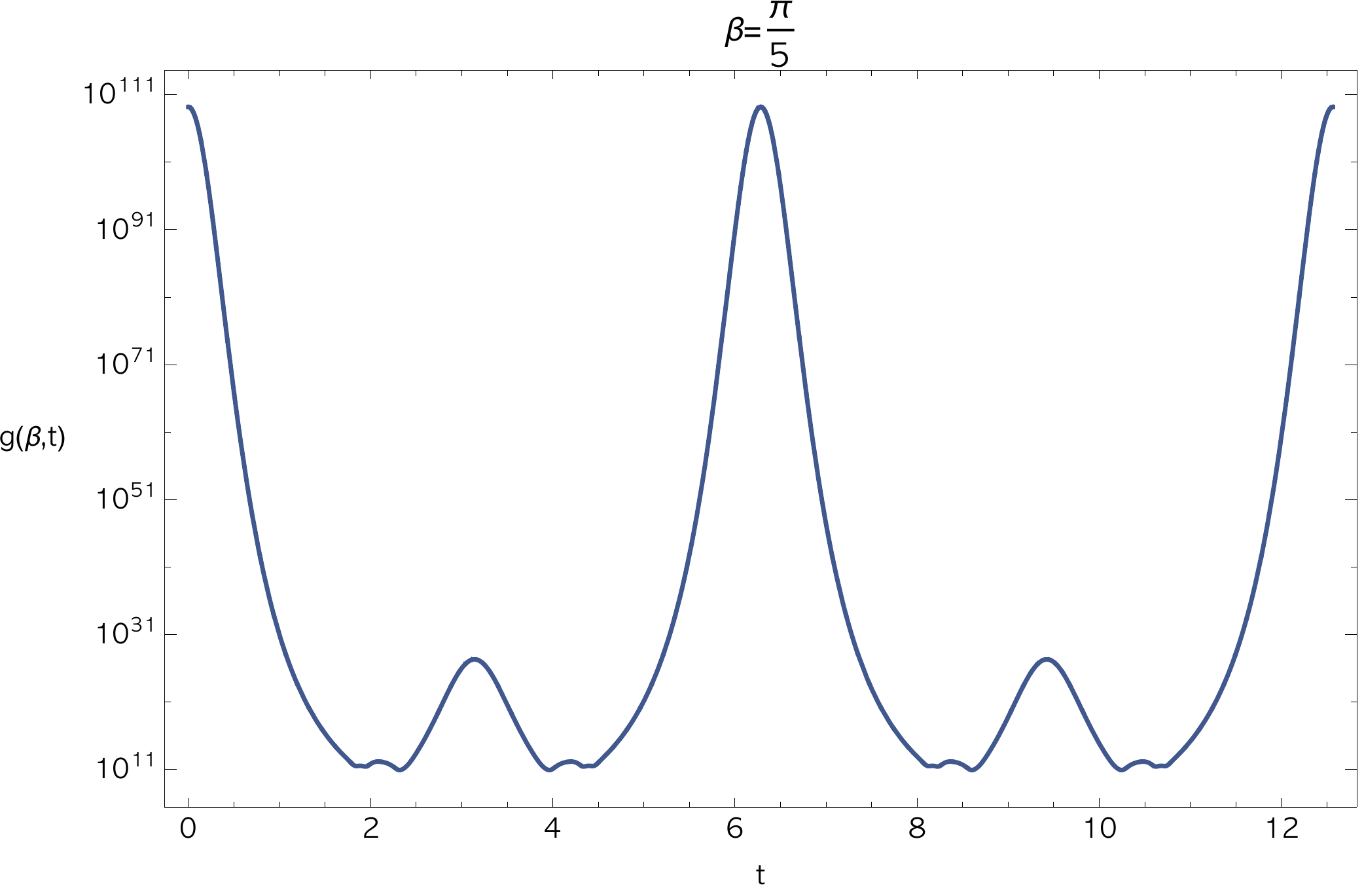}
	\includegraphics[width=0.24\textwidth]{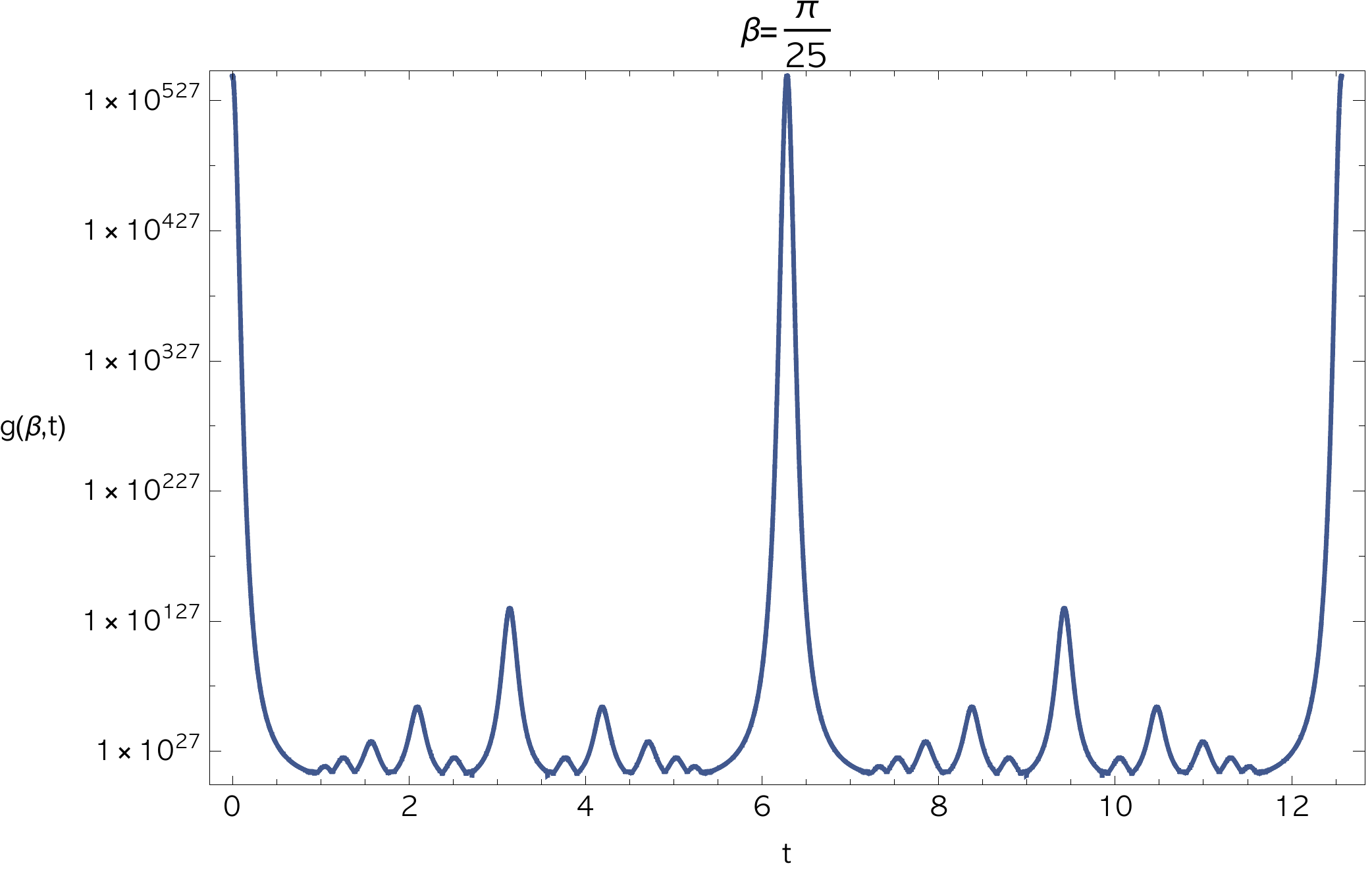}
      \caption{Here, in the top line, we display the behavior of our universal contribution, $g_{\star}(\beta,t)$ at various temperatures.
On the bottom line, for comparison, we display the spectral form factor for a sample modular invariant function $\psi_{2}(\tau)$.
As we increase temperature both are controlled by more and more saddles.}
\label{fig:gunivTemps}
\end{figure}

We can understand this from the $\Gamma_{\infty}\backslash SL(2;\mathbb{Z})$ tiling of the upper half plane, see Figure \ref{fig:funddom}.
As we increase temperature, we decrease $\mathrm{Im}(\tau)$, and intersect more and more fundamental domains.
Each such fundamental domain corresponds to a different modular image of the vacuum dominating.
At a given temperature, we can refine our identification of the universal contribution to the partition function,
\es{gnewuniv}{
Z_{\star}(\beta,t_{n/m})\equiv \chi_{0}(\gamma_{\star}(\tau_{n/m}))\bar{\chi}_{0}(\gamma_{\star}(\bar{\tau}_{n/m}))\,.
}
Here $\gamma_{\star}$ is the modular transformation that maximizes the vacuum character contribution at given temperature and time.\footnote{
Explicitly, given $\tau_{n/m},\bar{\tau}_{n/m}$ it is defined by
\begin{align}
  \gamma_* \equiv \underset{\gamma_{m,n}}{\mathrm{argmax}}
  |
  \chi_{0}(\gamma_{m,n}(\tau_{n/m}))
  \bar{\chi}_{0}(\gamma_{m,n}(\bar{\tau}_{n/m}))
  |^2 \ed
\end{align}}
At high temperatures, \eqref{gnewuniv} gives a complicated contribution to the partition function.
See Figure \ref{fig:gunivTemps} for an example.
At late times, it is easy to check that taking the decaying result, $\frac{e^{8 \pi ^2 k / \beta}}{t_n^{3}}$, for the spectral form factor, and replacing $t_n$ by an arbitrary time $t_{n/m}$, leads to a result that is always greater than or equal to $|Z_{\star}|^2$.

\subsection{Dip Time Estimate}
\label{sec:dip}

In this section we derive an upper bound on the time at which the spectral form factor of a generic chaotic CFT is expected to cross over to random matrix theory behavior.
We call this the dip time.
The derivation assumes that the universal contribution computed in previous sections correctly describes the late time behavior of the spectral form factor up to exponentially late times, right up to the dip time $t_d$.

The universal contribution, which we shall call the \textit{slope}, is bounded from above by
\begin{align}
  g_{\rm slope}(\beta,t) \sim \frac{e^{8\pi^2 k /\beta}}{t^s} \,,
  \label{gslope}
\end{align}
where $s=3$ for the vacuum character in the $\gamma_n$ frame, and $s=1$ for non-vacuum characters (where both $h,\bar{h}$ are non-zero).
While the result \eqref{gslope} was derived for the discrete times $t_n = 2\pi n$, as we saw in Section~\ref{sec:rathot} it provides an upper bound on the universal contribution and that will suffice for the purpose of deriving a bound.\footnote{
Notice that the universal contribution at non-integer times is exponentially smaller in $k$ than \eqref{gslope}.
Therefore, in practice we expect the random matrix theory contribution (the ramp) to `peak through' at non-integer times even before our estimate of the dip time. We thank Steve Shenker for pointing this out.
}

The decaying contribution \eqref{gslope} cannot be the full answer for a theory with a discrete spectrum at arbitrarily late times, because it violates the bound \eqref{gbound}.
Going to late times in the spectral form factor is equivalent to probing small energy differences in the spectrum.
At sufficiently late times we expect the properties of the spectrum at small energy differences (and therefore the behavior at very late times) to be goverened by random matrix theory \cite{Cotler2016}.
As described in Section~\ref{sec:rmt}, random matrix theory gives another universal contribution.
While this contribution is expected to have large fluctuations, on average its behavior is relatively simple.
Roughly speaking, it grows linearly in time until the plateau time $t_p$, beyond which it levels off at its asymptotic value which we shall denote $g_p$.

In this section we estimate the dip time $t_d$, which is the crossover time from the universal decay of \eqref{gslope} to the random matrix theory behavior.
We find that the ratio $t_p / t_d$ is exponentially large in $k$, which implies that there is a long period during which we expect the spectral form factor to grow linearly (on average) in a generic theory.

To get the late time behavior of the ramp and the plateau, recall that the thermodynamic partition function is given by the BTZ black hole partition function, $Z(\beta) = e^{8\pi^2 k / \beta}$.
The plateau height $g_p$ is bounded below by $Z(2\beta)$ (it can be pushed higher by degeneracies, which we ignore for now).
\begin{align}
  g_p \ge Z(2\beta) = e^{4\pi^2 k / \beta} \,.
\end{align}
The plateau time can be approximated by counting the available states at $2\beta$, so it is given by\footnote{
  The factor of 2 comes from the two terms in the exponent $e^{-\beta(E_n+E_m)}$ that appears in the sum over energy states.
  }
\begin{align}
  t_p \approx e^{S(2\beta)} = e^{8\pi^2 k / \beta} \,.
\end{align}
The ramp grows linearly in time, and should reach the plateau height at the plateau time.
The spectral form factor on the ramp is then given by
\begin{align}
  g_{\rm ramp}(t) = \frac{g_p t}{t_p} \ge e^{-4\pi^2 k/\beta} t \,.
\end{align}
The dip time $t_d$ is defined by $g_{\rm slope}(t_d) = g_{\rm ramp}(t_d)$, and is given by
\begin{align}
  t_d = \exp \left( \frac{12 \pi^2 k}{(1+s)\beta} \right)
\end{align}
For both the vacuum and matter contributions it is parametrically smaller than the plateau time:
\begin{align}
  \frac{t_p}{t_d} = \exp \left(
    \frac{2s-1}{s+1} \frac{4\pi^2 k}{\beta}
  \right) \,.
\end{align}
\subsection{Fine Spectral Probe}
\begin{figure}\label{fig:finestruc}
  \centering
	\includegraphics[width=0.45\textwidth]{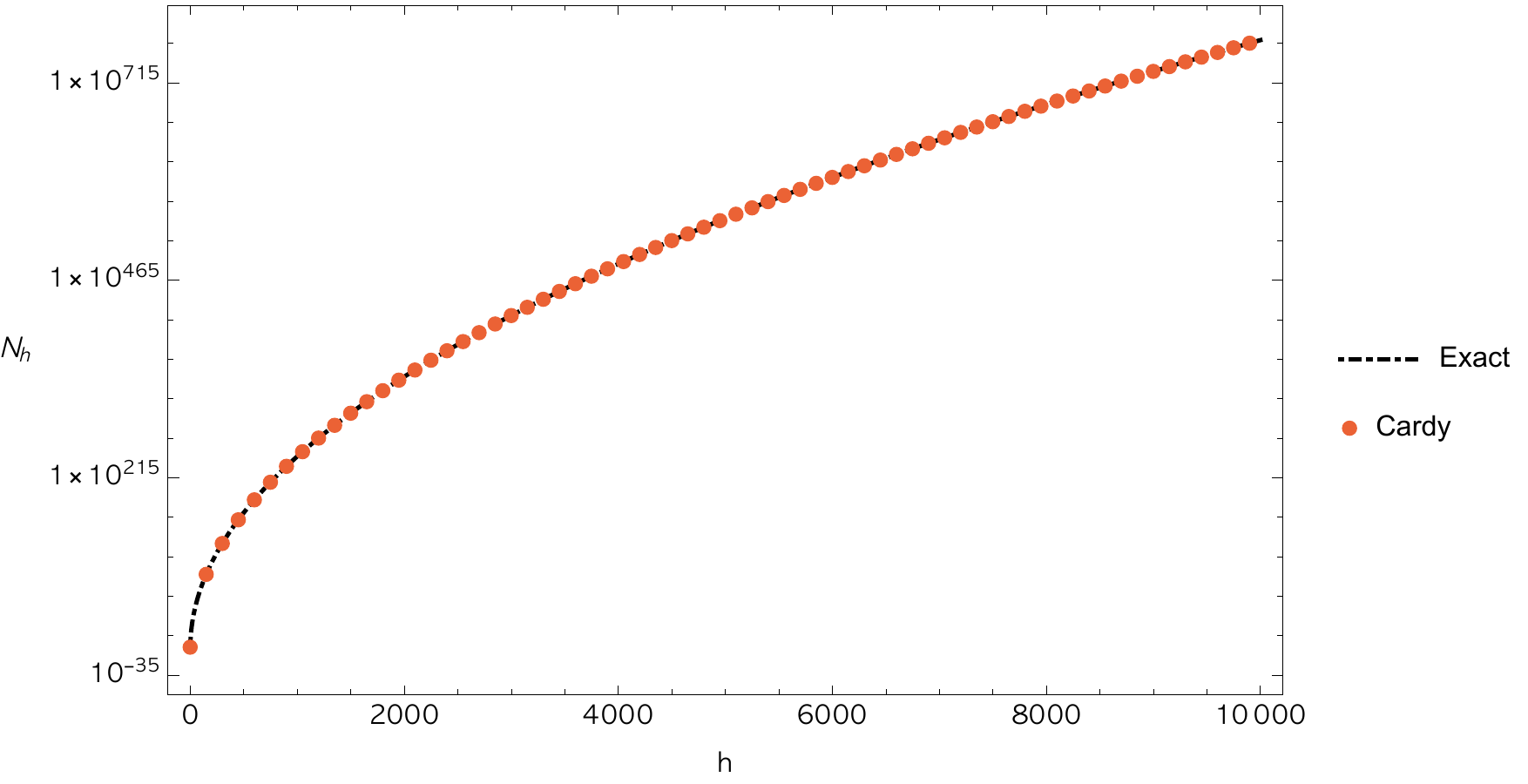}
	\includegraphics[width=0.45\textwidth]{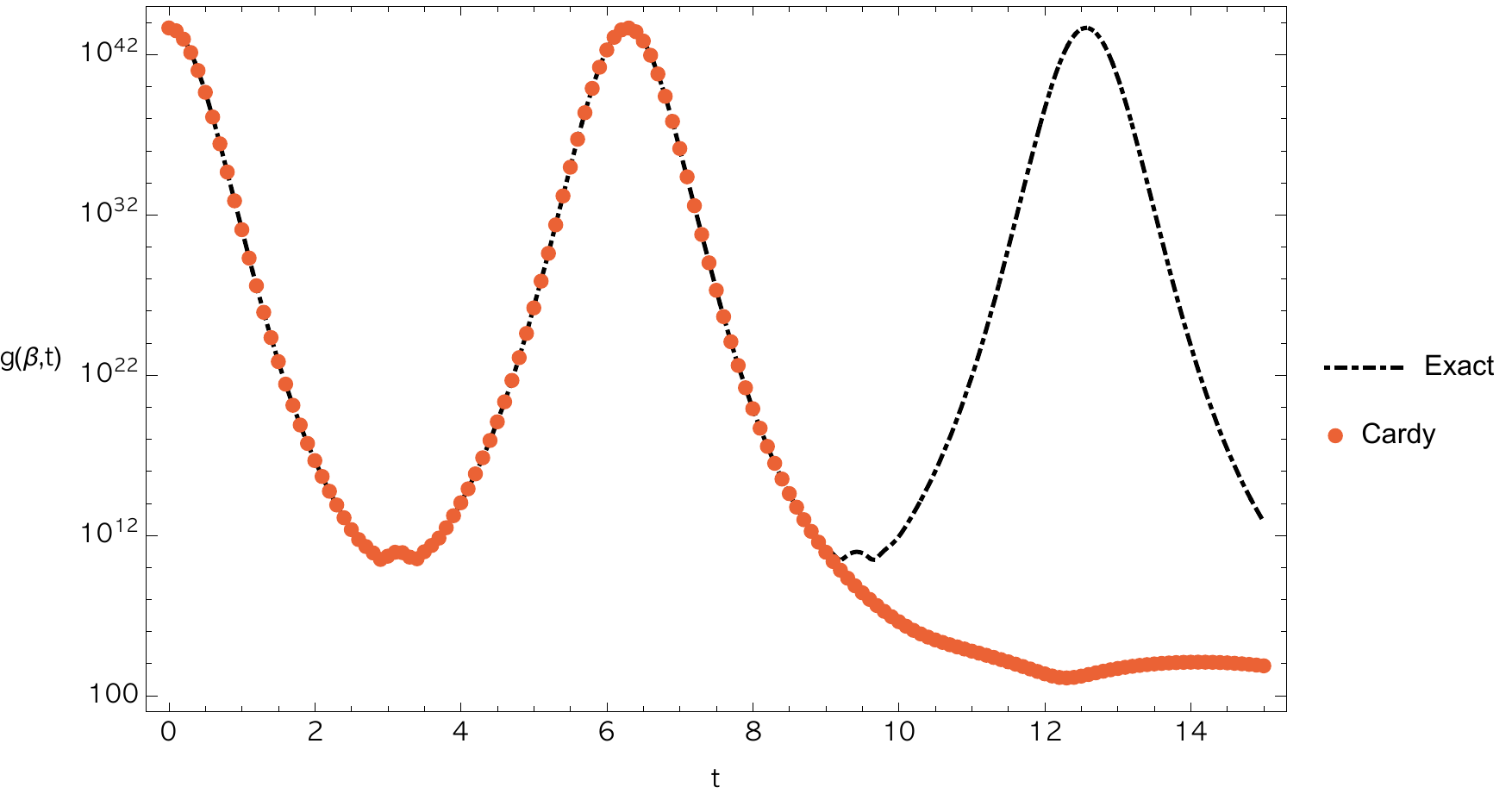}
      \caption{As an example here we plot the time dependence of an exact modular invariant function, $\psi_{2}$, and a function built out of a Cardy-like number of states at the same dimensions.
The left figure compares the exact spectrum in black with the Cardy spectrum in red, while the left figure shows the difference in the time dependence.}
\end{figure}

As we have seen discreteness of the spectrum in the original $SL(2,\mathbb{Z})$ frame is a necessary and sufficient condition for the partition function not to decay at late times.
However, modular invariance means that we should be able to present the partition function as a sum over states in any $SL(2;\mathbb{Z})$ frame.
\es{partfunc}{
Z(\tau,\bar{\tau})&=\sum_{h,\bar{h}}e^{2\pi i \left(h\gamma(\tau)-\bar{h}\gamma(\bar{\tau})\right)}\,.
}
In other frames, discreteness of the spectrum is not sufficient to guarantee the correct late time behavior, for instance a discrete set of states in the BTZ frame, may certainly decay.
Thus, the late time behavior probes slightly different features of the spectrum when viewed in each frame.
Of course, if we have a modular invariant spectrum these are all equivalent, but if one doesn't know a-priori that a given spectrum is modular invariant, the late time behavior in other frames provides a detailed probe.
To demonstrate this phenomenon, consider the time dependence depicted in Figure \ref{fig:finestruc}, where we compare the exact partition function, to the behavior of an approximate partition function built out of a discrete spectrum with exponentially small modifications to the degeneracies.
For long enough times, these two putative partition functions diverge despite the similarity in their spectra.
In this way, the time dependence in different frames probes detailed aspects of the CFT spectrum.

\section{Information Restoration in Integrable Theories}
\label{sec:int}

So far we have discussed information loss in chaotic CFTs. In Section~\ref{sec:uni} we have identified a decaying universal contribution to the spectral form factor, and commented on the expected late time behavior from random matrix theory.
In integrable theories we can say significantly more about the time dependence of the spectral form factor.\footnote{
  The same techniques can be applied to BPS subsectors of generic theories.
  }
Such theories are not chaotic and are not described by random matrix theories at small energy differences.
Therefore, their spectral form factors do not exhibit a dip, ramp, and plateau at late times.
Nevertheless, such theories do exhibit information loss at the level of individual Virasoro characters: Each Virasoro character still decays to zero at late times.
It is interesting to ask how information is restored in these simpler cases.

In this section we will answer this question for chiral CFTs.
The existence of chiral CFTs with large central charge that are dual to some form of semiclassical gravity is somewhat speculative \cite{Witten:2007kt, Li:2008dq, Gaiotto:2007xh, Gaberdiel:2007ve,Gaberdiel:2008xb, Benjamin:2015ria, Harrison:2016hbq, Benjamin:2016aww, Bae:2016yna}.
Here we will work under the assumption that such theories do exist, and that they have a sensible bulk interpretation (though the calculation itself will be done purely in field theory).

We will identify a set of modular transformations whose vacuum images are sufficient to restore information.
In generic non-chiral theories, the same set of transformations is responsible for the universal late time decay discussed in Section~\ref{sec:uni}.
In chiral theories, these transformations are enough to avoid the late time decay.

While we will focus on chiral theories, we note that much of what we say here also applies to holomorphic objects in general non-chiral theories, such as the elliptic genus which counts BPS states in theories with  $\mathcal{N}=(1,1)$ supersymmetry.

We now turn to a brief review of the properties of chiral CFTs.
In two spacetime dimensions, the vector representation of the Lorentz group is reducible into left-moving and right-moving representations.
Chiral conformal field theories are theories of purely left-moving degrees of freedom in Lorentzian signature, or purely holomorphic fields in Euclidean signature.
The symmetry algebra of these theories contains a single left-moving copy of the Virasoro algebra, and correspondingly a chiral CFT is labeled by a single central charge $c$.
Operators are labeled by a single conformal dimension, $h=\Delta=J$, where $J$ is the spin.

The torus partition functions of chiral CFTs can be written in a similar fashion to a generic $2d$ CFT.
\es{ChiralPart}{
  Z(\tau) &= \sum_{h\geq0}N_{h}q^{h-k}
  = \sum_{h\geq0} n_{h}\chi_{h}(\tau)
  \,,\qquad 
  k \equiv \frac{c}{24}\in\mathbb{Z}
  \,.
}
We again will be focusing on the case of modular invariant theories,
\es{modinvchi}{
  Z\left(\gamma(\tau)\right)&=Z\left(\tau\right)\,, \ \ \ \ \
  \gamma(\tau) = \frac{a\tau+b}{c\tau+d} \,, \ \ \ \ \
  \gamma \in\, SL(2;\mathbb{Z}) \,.
}
Modular invariant chiral CFTs are quite rigid.
First, $k$ and all conformal dimensions $h$ must be integers.
For this reason, the spectral form factor is periodic in time with an $\cO(1)$ period.
Second, the partition function is both modular invariant and meromorphic.
Such functions are uniquely determined by their poles and by the constant piece in the $q$ expansion (\ref{ChiralPart}) about $\tau=i\infty$.

As above, we will focus on sparse theories with $N_{h}\lesssim e^{2\pi h}$, for which the thermal partition function undergoes a sharp phase transition in temperature.
\es{Zchiralphase}{
\log Z(\beta)&=\left\{\begin{array}{ll}k\beta \,, & \beta>2\pi\\ \frac{4\pi ^2 k}{\beta } \,, & \beta<2\pi \end{array}\right. +\mathcal{O}(1)\,.
}
At high temperature the BTZ contribution dominates and is given by
\begin{align}
  Z_{\rm BTZ}(\tau) = \chi_0(-1/\tau) \,,
\end{align}
where $\tau = \frac{i\beta}{2\pi}$ as before.\footnote{
  We are calling this the `BTZ partition function' because it is dual to the contribution from the BTZ configuration in chiral gravity.
  See Appendix~\ref{app:bh} for details.
  }

We now analytically continue $\beta \to \beta + it$ as before, with the modular parameter given by
\begin{align}
  \tau = \frac{i\beta}{2\pi} - \frac{t}{2\pi} \ed
\end{align}
We consider the spectral form factor $g(\beta,t)=\left|Z(\beta+it)\right|^{2}$.
Just as in the non-chiral case, the BTZ contribution decays to zero at late times,
\begin{align}
  |Z_{\rm BTZ}(\tau)|^2 \sim 
  \frac{1}{t^{3}} \exp \left[ \frac{8\pi^{2}k\beta}{\beta^{2}+t^{2}} \right] \,.
\end{align}
We see that we have a phenomenon of information loss even in chiral theories.

It is now easy to see how information is restored.
The partition function is manifestly $2\pi$-periodic in time as a result of modular invariance, $Z(\tau) = Z(\tau + 1)$.
At time $t_n = 2\pi n$, $n \in \bZ$, the partition function is dominated by the modular image $\chi_0(\gamma_n(\tau_n))$ of the BTZ contribution.
This image is simply equal to $\gamma_0(-1/\tau_0)$ due to the periodicity.
As advertised, the modular transformation at time $t_n$ is the same one that gives the universal late time decay discussed in Section~\ref{sec:uni}.

\subsection{Saddle Point Expansion}

Our next goal is to describe, in bulk language, the mechanism by which information is restored.
The modular-invariant partition function includes contributions from $SL(2,\bZ)$ images of the vacuum character.
They are dual to a family of black holes in the bulk.
In this section we will explain that the partition function can be written as a sum over these saddle point contributions.
This description of the partition function is evocative of a bulk path integral.
In the next section we will discuss how information is restored in this saddle point expansion, and what this may teach us about the bulk.

As mentioned above, meromorphic modular invariant functions are entirely fixed by their poles and their constant term.
For a chiral CFT, this means that the full partition function,
\es{Zsplist}{
Z(\tau)&=\underbrace{\sum_{h=0}^{k}N_{h}q^{h-k}}_{Z_{L}(\tau)}+\underbrace{\sum_{h=k+1}^{\infty}N_{h}q^{h-k}}_{Z_{H}(\tau)}\,,
}
is fixed by the light spectrum --- those states with $h\leq k$.
Here the generating function for the light states is denoted by $Z_{L}$.

The way in which the spectrum of heavy states is fixed is relatively simple, and goes back to the work of Rademacher \cite{rademacher1937convergent, Rademacher:1968796}.\footnote{This mathematical structure is essentially the same for the generating functions of BPS states alluded to at the beginning of this section \cite{Dijkgraaf:2000fq}.}
We would like to complete $Z_{L}(\tau)$ into a fully modular invariant function.
One way to do this is to sum over the modular group, $SL(2;\mathbb{Z})$.
One generator, $\tau\rightarrow\tau+1$ acts trivially on $q$, so we only actually need to sum over $\Gamma_{\infty}\backslash SL(2;\mathbb{Z})$.
\es{radsum}{
Z(\tau)&=\sideset{}{'}\sum_{\gamma\in\Gamma_{\infty}\backslash SL(2;\mathbb{Z})}Z_{L}(\gamma(\tau))\,.
}
Here, the sum runs over the elements,
\es{gammadef}{
\Gamma_{\infty}\backslash SL(2;\mathbb{Z})\,=\,\left\{\gamma(\tau)\,=\,\frac{a\tau+b}{c\tau+d}:ad-bc=1, c=0,a=1 || 0\leq a<c\right\} \ec
}
which can be parameterized by the pair $(c,d)$ satisfying~ $\mathrm{gcd}(c,d)=1$.
The sum is primed to indicate that there is a regularization needed.
There is some freedom in how to regularize, but choices that preserve modular invariance can differ by at most an additive constant.\footnote{One simple way to regularize is to promote $Z(\tau)$ from a modular invariant function to a modular form of weight $w$,
$Z_{w}(\gamma(\tau))=(c\tau+d)^{w}Z_{w}(\tau)=\sum_{\gamma\in\Gamma_{\infty}\backslash SL(2;\mathbb{Z})}\frac{Z_{L}(\gamma(\tau))}{(c\tau+d)^{w}}\,.$
The partition function $Z(\tau)$ is then defined by analytic continuation.}

The sum takes on a particularly attractive meaning when thought of in the context of large $k$ CFTs dual to large radius gravity.
\es{sum}{
Z(\beta)&=Z_{L}(\beta)+Z_{L}(4\pi^{2}/\beta)+\sideset{}{'}\sum_{\substack{\Gamma_{\infty}\backslash SL(2;\mathbb{Z})\\c\geq1,d>0}}Z_{L}(\gamma_{c,d}(\tau))\Big |_{\tau=i\beta/2\pi}\\
&\approx e^{\beta k}+e^{\frac{4\pi^{2}k}{\beta}}+\sideset{}{'}\sum_{\substack{\Gamma_{\infty}\backslash SL(2;\mathbb{Z})\\c\geq1,d>0}}Z_{L}(\gamma_{c,d}(\tau))\Big |_{\tau=i\beta/2\pi}
}

It is tempting to identify this sum with the sum over bulk geometries.
In this description the first and second terms correspond to the vacuum and BTZ black hole respectively, and the remaining terms correspond to the subleading geometries $\mathcal{M}_{c,d}$ and their appropriate generalization for gravitational theories with matter.
As we review in Appendix~\ref{app:bh}, this can be made precise in the context of chiral gravity.

\subsection{Late Time Behavior in Saddle Point Expansion}

\begin{figure}\label{fig:chisad}
  \centering
    \includegraphics[width=0.8\textwidth]{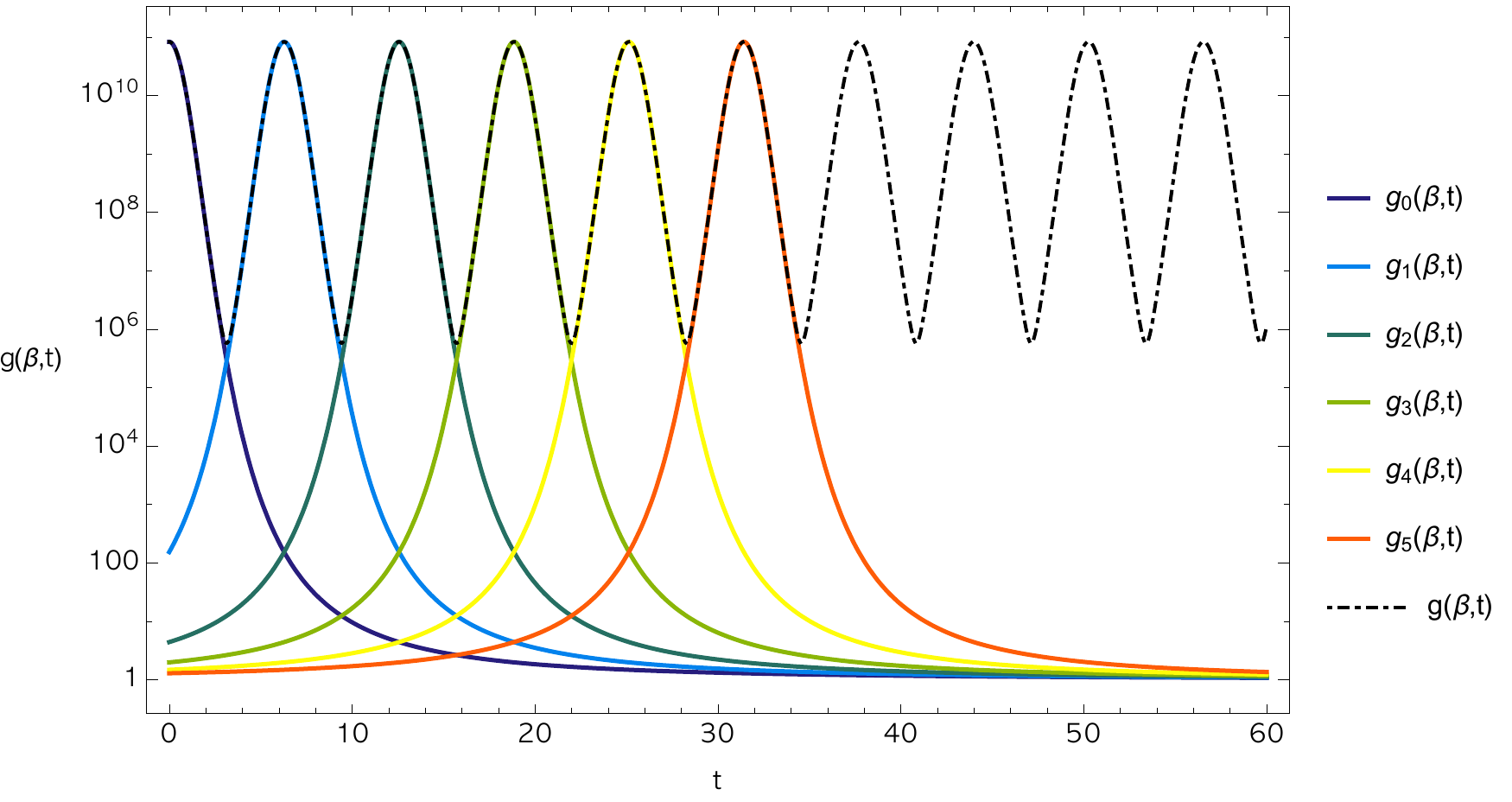}
    \caption{The spectral form factor $g(\beta,t)$ (dashed-dotted), and the contribution of six individual saddles $g_n(\beta,t)$, $n=0,\dots,5$ (solid lines). Each individual saddle $g_n$ is dominant around $t=t_n$ and exponentially sub-dominant at other times.
}
\end{figure}

Equipped with our expression of the partition function as an infinite sum over saddles, (\ref{radsum}), we can gain more insight into how the thermal partition function avoids late time decay.
Initially, at high temperatures, the partition function is well approximated by the BTZ contribution.
\es{Zinit}{
Z(\beta+it)&\approx Z_{\rm BTZ}(\beta+it)\,=\,e^{\frac{4\pi^{2}k}{\beta+it}}\,, \ \ \ \ \ 0<t\ll \beta \ed
}
This contribution, however, quickly begins to underestimate the partition function.
Focusing on times $t\approx t_{n}=2\pi n$ and taking $n>0$, the dominance of the BTZ saddle is eclipsed by the appropriate saddle, labeled by $(c,d)=(1,n)$.
\es{dsad}{
Z(\beta+it)&\approx Z(\gamma_{n}(\tau))\Big |_{\tau=\frac{i(\beta+it)}{2\pi}} \,\approx\,e^{\frac{4\pi^{2}k}{\beta+i(t-2\pi n)}}\,, \ \ \ \ \ 0<t\ll\beta\,.
}
For each integer $n$ the given saddle goes from subdominant to dominant and then exponentially decays again.
Only by summing over this infinite class of saddles do we get a partition function that exhibits the appropriate, non-decaying behavior, see Figure \ref{fig:chisad}.

For non-integer time, we again have the spaghetti like behavior of section (\ref{sec:rathot}).
For each time $t=n/m$ there is a phase transition such that for all $\beta<\beta_{m,n}$ we are dominated by the $(m,n)$ saddle.
In this way, reproducing the correct late time behavior at all temperatures depends crucially on including the appropriate set of saddles.

\subsection{Discretizing The Spectrum}

Throughout this paper we have emphasized the connection between the late time behavior of the spectral form factor and the discrete nature of the spectrum.
In this section we review how the naively smooth spectral density is rendered discrete by the $SL(2;\mathbb{Z})$ saddle point expansion.
Including a large but finite number of saddles in the expansion yields a smooth density of states with sharp peaks around the locations of the underlying states, while including all saddles leads to a fully discrete density of states (c.f. eq. \eqref{densitysimp}).

To be concrete, we will study weight $w$ modular forms $\psi_{n;w}$, with polar part consisting of a single pole of weight $n$.
\es{psiw}{
\psi_{n;w}(\tau)&\equiv\frac{1}{q^{n}}+\cO(q)\,.
}
They have the following property under modular transformation.
\es{psiwtrans}{
\psi_{n;w}\left(\gamma(\tau)\right)&=(c\tau+d)^{w}\psi_{n;w}(\tau)\,.
}
To make contact with the previous discussion, the functions $\psi_{n;0}$ can be used as a basis for constructing a partition function.
Strictly speaking, the manipulations we present are only valid for $w>1$, but we may think of introducing $w$ as a regulator.\footnote{
A special case of this is the differential regularization advocated in \cite{Maloney:2007ud, Manschot:2007ha}.}
The final results can be analytically continued to $w=0$.
They match careful computations performed in the $w<1$ regime with a subtraction based scheme \cite{rademacher1937convergent, Rademacher:1968796}.
For $w=0$ the only holomorphic modular function is a constant, and so any scheme that preserves modular invariance is guaranteed to reproduce the same modular function, up to a constant.
This constant may be important for understanding whether theories of pure $3d$ gravity exist \cite{Gaberdiel:2008xb, Benjamin:2016aww}, but will not effect our discussion here.

Given any $T$ invariant function, $f(\tau+1)=f(\tau)$, we may write,
\es{F_{w}}{
F_{w}(\tau)&=\sum_{\gamma\in\Gamma_{\infty}\backslash SL(2;\mathbb{Z})}\frac{1}{(c \tau+d)^{w}}f(\gamma(\tau))\,.
}
To see how $F_{w}(\tau)$ transforms, we apply an element of $SL(2;\mathbb{Z})$.
\es{F_{w}trans}{
F_{w}(\gamma(\tau))&=\sum_{\gamma^{\prime}\in\Gamma_{\infty}\backslash SL(2;\mathbb{Z})}\frac{1}{(c^{\prime} \gamma(\tau)+d^{\prime})^{w}}f(\gamma^{\prime}(\gamma(\tau)))\\
&=\sum_{\gamma^{\prime\prime}\in\Gamma_{\infty}\backslash SL(2;\mathbb{Z})}\left(\frac{c\tau+d}{c^{\prime\prime} \tau+d^{\prime\prime}}\right)^{w}f(\gamma^{\prime\prime}(\tau))\\
&=(c\tau+d)^{w}F_{w}(\tau)\,.
}
The one subtlety in the above argument is working with the cosets, $\Gamma_{\infty}\backslash SL(2;\mathbb{Z})$ rather then the full group, but as $f$ is $T$ invariant, and $\{c,d\}$ do not change when acting with $T$ on the left, we are free to work in the coset space.

We are interested in the special case,
\es{psiwsum}{
\psi_{n;w}(\tau)&=\sum_{\gamma\in\Gamma_{\infty}\backslash SL(2;\mathbb{Z})}\frac{e^{-2\pi i n \gamma(\tau)}}{(c \tau+d)^{w}}\\
&=\frac{1}{q^{n}}+\sum_{m\geq0}N_{m}^{(n;w)}q^{m}\,.
}
In terms of a real inverse temperature, we can write
\es{psibeta}{
\psi_{n;w}(\beta)&= e^{\beta n}+\int_{0}^{\infty}d\Delta\, \rho_{(n;w)}(\Delta)e^{-\beta \Delta}\,,
}
and perform an inverse Laplace transform to read off the density of states.
The term involving the density of states can be written explicitly as,
\es{densitysum}{
\int_{0}^{\infty}d\Delta\, \rho_{(n;w)}(\Delta)e^{-\beta \Delta}&=\sum_{\gamma\in\left(\Gamma_{\infty}\backslash SL(2;\mathbb{Z})\right)^{*}}\frac{e^{-2\pi i n \gamma(\tau)}}{(c \tau+d)^{w}}\,,
}
where the $*$ indicates that we have dropped the identity contribution from the modular sum.
Performing the inverse Laplace transform gives,
\es{density}{
\rho_{(n;w)}(\Delta)&=\frac{1}{2\pi i}\sum_{\gamma\in\left(\Gamma_{\infty}\backslash SL(2;\mathbb{Z})\right)^{*}}\int_{\epsilon-i\infty}^{\epsilon+i\infty}d\beta\,e^{\beta\Delta}\frac{e^{-2\pi i n \gamma(\tau)}}{(c \tau+d)^{w}}\,.
}
It is useful to organize the sum over $\Gamma_{\infty}\backslash SL(2;\mathbb{Z})$ as a double sum first over $\Gamma_{\infty}\backslash SL(2;\mathbb{Z})/\Gamma_{\infty}$, and a sum over right action by $T^{\ell}$.
Then, by using the identity,
\es{singleterm}{
\frac{1}{2\pi i}\int_{\epsilon-i\infty}^{\epsilon+i\infty}d\beta\,e^{\beta\Delta}\frac{e^{-2\pi i n \gamma(\tau)}}{(c \tau+d)^{w}}&=-2\pi\left(-i \sqrt{\frac{\Delta }{n}}\right)^{w-1}\frac{e^ {2\pi i(\Delta\frac{d}{c}-n\frac{a}{c})}}{c}I_{w-1}\left(\frac{4\pi}{c}\sqrt{n\Delta}\right)\,,
}
we can write the density as,
\es{densitysimp}{
\rho_{(n;w)}(\Delta)&=-2\pi\sum_{d=0:\,(c,d)=1}^{c-1}\sum_{\ell=-\infty}^{\infty} e^{2\pi i \Delta \ell}\left(-i \sqrt{\frac{\Delta }{n}}\right)^{w-1}\frac{e^ {2\pi i(\Delta\frac{d}{c}-n\frac{a}{c})}}{c}I_{w-1}\left(\frac{4\pi}{c}\sqrt{n\Delta}\right)\\
&=-2\pi\sum_{d=0:\,(c,d)=1}^{c-1}\sum_{s=-\infty}^{\infty} \delta(\Delta-s)\left(-i \sqrt{\frac{\Delta }{n}}\right)^{w-1}\frac{e^ {2\pi i(\Delta\frac{d}{c}-n\frac{a}{c})}}{c}I_{w-1}\left(\frac{4\pi}{c}\sqrt{n\Delta}\right)\,.
}
The delta function in the last line is exactly the discreetness of the spectrum we were after.
Notice that including a finite number of saddles, by placing a cutoff on $|\ell|$, leads to a smooth density of states that becomes progressively sharper around the discrete states as we increase the cutoff.
Put differently, by including an increasing number of saddles in the expansion we can witness the discreteness of the spectrum emerge out of the smooth density.
\section{Discussion}

In this paper we have examined the time dependence of the partition function in two-dimensional conformal field theories.
We identified a universal contribution which decays slowly in time. By apealing to the late time behavior of random matrix theory we were able to conjecture a dip time, where we expect the crossover to RMT to set in.
In integrable models, in particular chiral conformal field theories, we were able to identify an infinite set of saddle point contributions to the partition function, corresponding to black holes in the bulk, which serve to restore information for all time.
All of these discussions, however, leave open many avenues of future inquiry.

One important question is when do the correction to the Cardy formula \eqref{cardyChi} describing the density of characters become important enough to affect the late time behavior.
In theories with sufficiently sparse spectra, we expect such corrections to be responsible for the late time transition to random matrix theory behavior.
They may also affect the universal decay worked out in Section~\ref{sec:uni} before the dip time.
A possible starting point for investigating these questions is to include non-vacuum states on the left-hand side of \eqref{cardyChi}.

An important assumption we use was sparsity of the light spectrum in gravitational theories. An obvious question is how the notion of sparsity imposed here connects to other such criteria one may wish to impose for a conformal field theory dual to gravity. For instance those coming from requiring a Hawking-Page phase transition, appropriate behavior of R\'enyi entropies, saturation of Lyapunov bounds, or from demanding a bulk point singularity \cite{Hartman:2014oaa,Haehl:2014yla,Belin:2014fna,Benjamin:2015hsa,Benjamin:2015vkc,Benjamin:2016pil,Hartman:2013mia,Perlmutter:2016pkf,Roberts:2014ifa,Heemskerk:2009pn}. 

We have mentioned that the discussion of information loss in integrable theories can in principle be applied to counts of BPS states in generic supersymmetric theories. It would be interesting to study this in detail. It would be especially interesting if one can leverage information about how the BPS spectrum solves its information paradox to make statements about the full supersymmetric theory.

As we investigate the analytically continued partition function at higher and higher temperature, it's time dependance becomes very featured, see Figure~\ref{fig:gunivTemps}. For our universal contribution, as well as for chiral CFTs, there are spikes that occur at regular, rational times. An ambitious question is whether there is an experimental observable (perhaps considering a two point function rather than a partition function) which might be able to detect these rational spikes for experimentally realizable 1+1d systems.

\section*{Acknowledgements}

We thank Nathan Benjamin, A. Liam Fitzpatrick, Shamit Kachru, Jared Kaplan, Alex Maloney, Eric Perlmutter, and Mukund Rangamani for useful discussions.
We thank Stephen H. Shenker for useful discussions, insightful suggestions, and encouragement.
We thank Solomon Endlich and Masanori Hanada for tasty treats during the completion of this work.

G. G. is supported by a grant from the John Templeton Foundation. The opinions expressed in this publication are those of the authors and do not necessarily reflect the views of the John Templeton Foundation.

\appendix

\section{Euclidean Black Holes in AdS$_{3}$}\label{app:bh}
In this appendix we review some basic facts about thermal solutions to the vacuum Einstein equations.
The gravitational action in three dimensional negatively curved space is,
\es{EHact}{
S&=\frac{1}{16\pi G}\int\sqrt{g}d^{3}x\left(R+\frac{2}{\ell^{2}}\right)+\frac{1}{8\pi G}\int d^{2}x\sqrt{\gamma}\left(K+\frac{1}{\ell}\right)
}
with equations of motion,
\es{EHeom}{
G_{\mu\nu}^{\ell}&=R_{\mu\nu}-\frac{1}{2}Rg_{\mu\nu}-\frac{1}{\ell^{2}} g_{\mu\nu}\,=\,0
}
We are interested in thermal, finite volume, asymptotically AdS$_{3}$ solutions, that is solutions whose conformal boundary is a torus.
The most familiar such example is thermal AdS$_{3}$.
The metric is given by,
\es{AdS3met}{
ds^{2}&=(1+r^{2})dt^{2}+\frac{dr^{2}}{1+r^{2}}+r^{2}d\phi^{2}\,.
}
Here, $\phi=\phi+2\pi$ is an angular coordinate, and we identify $t=t+\beta$ for thermal AdS$_{3}$ with inverse temperature $\beta$.
The coordinates $t$ and $\phi$ parameterize the boundary torus, which is filled in by the radial coordinate, $r$.
the $\phi$ cycle is contractable.

Another familiar finite temperature solution is the BTZ black hole \cite{Banados:1992wn}.
This can be represented by the metric,
\es{BTZmet}{
ds^{2}&=(r^{2}-r_{+}^{2})dt^{2}+\frac{dr^{2}}{r^{2}-r_{+}^{2}}+r^{2}d\phi^{2}\,,
}
where $r\geq r_{+}$, $\phi$ is again periodic, $\phi=\phi+2\pi$, and $t$ is periodic, with periodicity set by ensuring the black hole is non-singular at the horizon, $t=t+2\pi/r_{+}$.
At the horizon, the $t$ cycle shrinks to zero size, while the $\phi$ cycle does not.
So the role of $\phi$ and $t$ have switched in terms of which cycle is contractable.

The AdS$_{3}$ and BTZ metrics look different, and have a different choice for which cycle is contractable, but in fact, the Euclidean BTZ black hole at temperature, $\beta=2\pi/r_{+}$ is diffeomorphic to thermal AdS$_{3}$, at $\beta^{\prime}=2\pi/\beta=r_{+}$.
To see this, we simply define the new coordinates,
\es{newchords}{
\phi^{\prime}&=r_{+}t\\
t^{\prime}&=-r_{+}\phi\\
r^{\prime}&=\sqrt{\left(\frac{r}{r_{+}}\right)^{2}-1}\,,
}
With this, the periodicities are, $\phi^{\prime}=\phi^{\prime}+2\pi$ and $t^{\prime}=t^{\prime}+2\pi r_{+}$.
The $\phi^{\prime}$ cycle is contractable, and the metric is the usual AdS$_{3}$ metric, (\ref{AdS3met}), with primed coordinates.

\subsection{$SL(2;\mathbb{Z})$ Black Holes}

The above story can be generalized in a number of ways, but the basic picture remains the same.
Firstly we can consider solutions at finite temperature and finite chemical potential.
For thermal AdS$_{3}$ at finite chemical potential we keep the metric (\ref{AdS3met}), but impose the more general identification,
\es{compid}{
z\equiv\phi+it&=z+2\pi=z+2\pi \tau\,.
}
Here we have defined the complex modulus, $\tau=(\mu+i\beta)/2\pi$.

We also have spinning BTZ black holes with finite temperature and chemical potential.
The metric,
\es{spinningBH}{
ds^{2}=\frac{(r^{2}-r_{+}^{2})(r^{2}-r_{-}^{2})}{r^{2}}dt^{2}+\frac{r^{2}}{(r^{2}-r_{+}^{2})(r^{2}-r_{-}^{2})}dr^{2}+r^{2}(d\phi+i\frac{r_{+}r_{-}}{r^{2}}dt)^{2}
}
is only non-singular if $z=z+2\pi m +2\pi n \tau$, with $\tau=i/(r_{+}+r_{-})$.

Again this euclidean black hole is diffeomorphic to pure thermal AdS$_{3}$ with chemical potential.
This can be seen directly, by defining,
\es{newchordsspin}{
\phi^{\prime}&=r_{+}t-ir_{-}\phi\\
t^{\prime}&=-r_{+}\phi-ir_{-}t\\
r^{\prime}&=\sqrt{\frac{r^{2}-r_{+}^{2}}{r_{+}^{2}-r_{-}^{2}}}\,.
}
This gives the AdS$_{3}$ metric in the primed variables, with the identification, $z=z+2\pi m+2\pi n \tau^{\prime}$, with $\tau^{\prime}=-1/\tau=i(r_{+}+r_{-})$.

In this way we can view the BTZ black hole at $\tau$ as AdS$_{3}$ at $-1/\tau$.
This is possible as the transformation $\tau\rightarrow-1/\tau$ is part of the modular group, $SL(2;\mathbb{Z})$, which preserves the boundary torus.

More generally, at a fixed temperature and chemical potential, we have an $SL(2;\mathbb{Z})$ family of black holes.
These were originally introduced in \cite{Maldacena:1998bw} and have been discussed extensively, for instance \cite{Dijkgraaf:2000fq, Maloney:2007ud}.

These black holes are all diffeomorphic to AdS$_{3}$, but with modulous $\gamma(\tau)=\frac{a\tau+b}{c\tau+d}$, for $\gamma\in SL(2;\mathbb{Z})$.
We thus naively have a family of solutions labeled by four integers, subject to the constraint, $ad-bc=1$.
However, the definition of $\tau$ in the quotient (\ref{compid}) is slightly redundant, $\tau\sim\tau+1$, and so the inequivalent configurations are really labeled by elements of $\Gamma_{\infty}\backslash SL(2;\mathbb{Z})$, ie relatively prime integers $(c,d)$, and are denoted $\mathcal{M}_{c,d}$.
\subsection{Black Hole Partition Function}
We are interested in evaluating the gravitational partition function around the classical saddles, $\mathcal{M}_{c,d}$.
As each is diffeomorphic to AdS$_{3}$, it is sufficient to evaluate the partition function around the metric (\ref{AdS3met}).

At the classical level, this involves evaluating the action (\ref{EHact}) on the classical solution, and gives,
\es{acteval}{
\log Z_{\rm vac} &\sim -2\pi i k(\tau-\bar{\tau})
}
Here we have used the identification, $c=24k=\frac{3\ell}{2G}$.

Either by using the correspondence with the dual 2d CFT, or by explicit computation \cite{Giombi:2008vd} the full quantum partition function around the classical saddle can be evaluated, giving the one loop exact result,
\es{chareval}{
\log Z_{\rm vac} &= -2\pi i (k-1/24)(\tau-\bar{\tau})+\log\left(|1-q|^{2}\right)-\log\left(|\eta(\tau)|^{2}\right)\,.
}

In evaluating these expressions, we have used the Einstein-Hilbert action, (\ref{EHact}).
We will also be interested in the story for ciral gravity \cite{Li:2008dq, Strominger:2008dp, Maloney:2009ck}, for which the action is modified with a gravitational Chern-Simons term.
\es{chiact}{
S_{\chi}&=\frac{1}{16\pi G}\int d^{3}x\sqrt{-g}\left[R+2-\frac{1}{2}\epsilon^{\mu\nu\rho}\Gamma_{\mu\gamma}^{\kappa}\left(\partial_{\nu}\Gamma_{\kappa\rho}^{\gamma}+\frac{2}{3}\Gamma_{\nu\delta}^{\gamma}\Gamma^{\delta}_{\rho\kappa}\right)\right]+S_{bdy}\,,
}
where the appropriate boundary term was discussed in \cite{Kraus:2006nb}. The equations of motion are given by,
\es{chieom}{
G_{\mu\nu}^{\ell}+\mathcal{C}_{\mu\nu}&=0\,,
}
with $\mathcal{C}$, the Cotton tensor.
\es{GCdef}{
\mathcal{C}_{\mu\nu}&=\epsilon_{\mu}^{\nu\rho}\nabla_{\nu}\left(R_{\rho\nu}-\frac{1}{4}g_{\rho\nu}R\right)
}
As $G^{\ell}_{\mu\nu}=0\rightarrow\mathcal{C}_{\mu\nu}=0$ any solution to the vacuum Einstein equations is a solution to chiral gravity.\footnote{As far as is known, the converse is also true \cite{Maloney:2009ck}.} In particular the metrics, $\mathcal{M}_{c,d}$ are still classical solutions and we can again ask what their contribution to the partition function yields.

At the classical level, we must now evaluate the action (\ref{chiact}) on $\mathcal{M}_{c,d}$.
This gives,
\es{chicont}{
\log Z_{\rm vac} &\sim -2\pi i k\tau\,.
}
which again is enhanced to the full character by one-loop corrections.
Here, in mapping from the gravitational result to the cft, we have used the fact that the asymptotic symmetry algebra of chiral gravity consists of a single chiral Virasoro algebra with
\es{cchiral}{
c\,=\,24k\,=\,\frac{3\ell}{G}\,.
}

\bibliographystyle{utphys}
\bibliography{refs}

\end{document}